\documentclass[prb,aps,superscriptaddress,floatfix,tightenlines,showpacs,twocolumn]{revtex4-2}
\usepackage{bm,amsmath,amssymb} 
\usepackage{graphicx}
\graphicspath{{FIG/}}
\usepackage{dcolumn}   
\usepackage{ulem}
\usepackage{url}
\usepackage{layout}
\usepackage{epsfig}
\usepackage{booktabs}
\usepackage{color}
\usepackage{float}
\usepackage[hyperindex,breaklinks]{hyperref}
\usepackage{comment}

\definecolor{BLACK}{gray}{0}
\definecolor{WHITE}{gray}{1}
\definecolor{RED}{rgb}{1,0,0}
\definecolor{GREEN}{rgb}{0,1,0}
\definecolor{BLUE}{rgb}{0,0,1}
\definecolor{CYAN}{cmyk}{1,0,0,0}
\definecolor{MAGENTA}{cmyk}{0,1,0,0}
\definecolor{YELLOW}{cmyk}{0,0,1,0}
\definecolor{armygreen}{rgb}{0.55, 0.73, 0.0}

\begin{document}

\title{Multicriticality of Two-dimensional Class D Disordered Topological Superconductors}

\author{Tong Wang}
\affiliation{International Center for Quantum Materials, School of Physics, Peking University, Beijing 100871, China}

\author{Zhiming Pan}
\affiliation{International Center for Quantum Materials, School of Physics, Peking University, Beijing 100871, China}

\author{Tomi Ohtsuki}
\affiliation{Physics Division, Sophia University, Chiyoda-ku, Tokyo 102-8554, Japan}%

\author{Ilya A. Gruzberg}
\affiliation{Department of Physics, Ohio State University, 191W. Woodruff Ave, Columbus OH, 43210, United States of America}%

\author{Ryuichi Shindou}
\email{rshindou@pku.edu.cn}
\affiliation{International Center for Quantum Materials, School of Physics, Peking University, Beijing 100871, China}%
\date{\today}

\begin{abstract}
A generic two-dimensional disordered topological superconductor in symmetry class D exhibits rich phenomenology and multiple phases: diffusive thermal metal (DTM), Anderson insulator (AI), and thermal quantum Hall (TQH) phase (a topological superconductor). We numerically investigate the phase diagram of a lattice model of such class D superconductor, specifically focusing on transitions between the phases and the associated universal critical behaviors. We confirm the existence of a tricritical point and its repulsive nature at the point on the phase diagram where the three phases meet. We characterize the critical behaviors at various critical points and the tricritical point using numerical evaluation of the localization length, the conductance (or conductivity), and the density of states. We conclude that the two metal-insulator transitions (DTM-TQH and DTM-AI) belong to the same universality class, whereas the tricritical point (TCP) represents a distinct universality class.
\end{abstract}


\maketitle

\section{Introduction}
\label{sec:introduction}

Low-energy quasiparticle fermionic excitations in some unconventional superconductors \cite{Read2000paired} and quantum magnets \cite{kitaev06} have an unusual property of being their own antiparticles. These real or Majorana fermions attracted much attention in condensed matter experiments in recent years \cite{he17, kayyalha20, kasahara18, wang18}. Subsequent studies stressed the importance of quenched disorder in experiments \cite{huang18, lian18, knolle19, yamada20}.

A canonical disordered system with low-energy Majorana quasiparticles is a two-dimensional (2D) disordered topological superconductor in symmetry class D \cite{altland1997nonstandard} modelled by a mean-field Bogoliubov-de Gennes Hamiltonian. The physics of Anderson localization and the thermal transport of quasiparticles in class D systems is extremely rich. Generically, such systems exhibit phase diagrams that encompass (thermal) Anderson insulators (AI), thermal quantum Hall (TQH) phases, or topological superconductors, and an enigmatic diffusive thermal metal (DTM) or `Majorana metal' phase  \cite{Senthil2000prb, Bocquet2000, Chalker2001prb, Wimmer2010prl, Laumann2012prb, Yoshioka18, Fulga-Temperature-2020}. In addition to metal-insulator transitions (MIT) and thermal quantum Hall transitions, class D systems may have tricritical points where all the three phases meet.

Another 2D disordered system in class D with Majorana particles results from fermionization of the random-bond Ising model (RBIM) in 2D \cite{Dotsenko1983, ChoFisher1997, Read2000absence, Gruzberg-Random-bond-2001, MerzChalker2002, Merz-Negative-2002, Mildenberger-Griffiths-2006}. The phase diagram of the RBIM does not contain a metallic phase \cite{Read2000absence}, but has an intriguing multicritical Nishimori point.

Both disordered superconductors in class D and the RBIM can be reformulated as  network models \cite{ChoFisher1997, Mildenberger-Griffiths-2006},
which are convenient for numerical simulations and have been extensively
studied \cite{ChoFisher1997, Chalker2001prb, Mildenberger-Griffiths-2006, Mildenberger2007, Evers2008RMP, Kagalovsky2008prl, Kagalovsky2010prb, Medvedyeva2010prb, mkhitaryan11,Lian2018prb}. However, many properties of the phases and phase transitions in class D disordered superconductors remain elusive. In particular, the nature and even the position of the tricritical point have remained unknown. Several possibilities were proposed regarding this point \cite{Mildenberger2007}, but few definitive conclusions have been made \cite{Kagalovsky2010prb, Medvedyeva2010prb}.

In this paper, we study a class D disordered $p_x+{\rm i}p_y$ superconductor
described by a tight-binding Bogoliubov-de Gennes Hamiltonian on a square lattice:
\begin{align}
	\mathcal{H}/2 &= \sum_{\bm j} (\varepsilon_{\bm j} +\mu) c_{\bm j}^\dagger c_{\bm j}^{\vphantom{\dagger}}
	+ \sum_{\bm j} \sum_{\nu=x,y} t_\nu \big[ c_{\bm j+\bm{e}_\nu}^\dagger  c_{\bm j}^{\vphantom{\dagger}}
	+ \mathrm{h.c.} \big] \nonumber \\
	& \quad +\Delta \sum_{\bm j} \big[\mathrm{i} c_{\bm j+\bm{e}_x}^\dagger
	c^{\dagger}_{\bm j} +  c_{\bm j+\bm{e}_y}^\dagger c^{\dagger}_{\bm j}
	+ \mathrm{h.c.} \big].
	\label{bdg}
\end{align}
Here $\mu$, $t_{\nu}$, and $\Delta$ are the chemical potential, the
nearest-neighbor hopping amplitudes, and the $p$-wave superconducting pairing amplitude, respectively. We parametrize the hopping amplitudes in two directions as $t_x = (1 - \alpha)t$, $t_y = (1 + \alpha)t$
with $0\le |\alpha|<1$. The system is isotropic at $\alpha= 0$ and anisotropic otherwise. Quenched disorder is represented by random on-site energies $\varepsilon_{\bm j}$ drawn from a certain distribution, and $\bm j\equiv (j_x,j_y)$ labels the lattice sites.
The Hamiltonian Eq.~\eqref{bdg} 
describes spinless fermions and breaks time-reversal symmetry,
thus belonging to class D \cite{altland1997nonstandard}.

Below, we give a comprehensive numerical characterization of different phases and phase transitions in this model, including a discussion of the nature of criticality at the tricritical points.

\begin{figure*}
\centering
\includegraphics[width=6.5in]{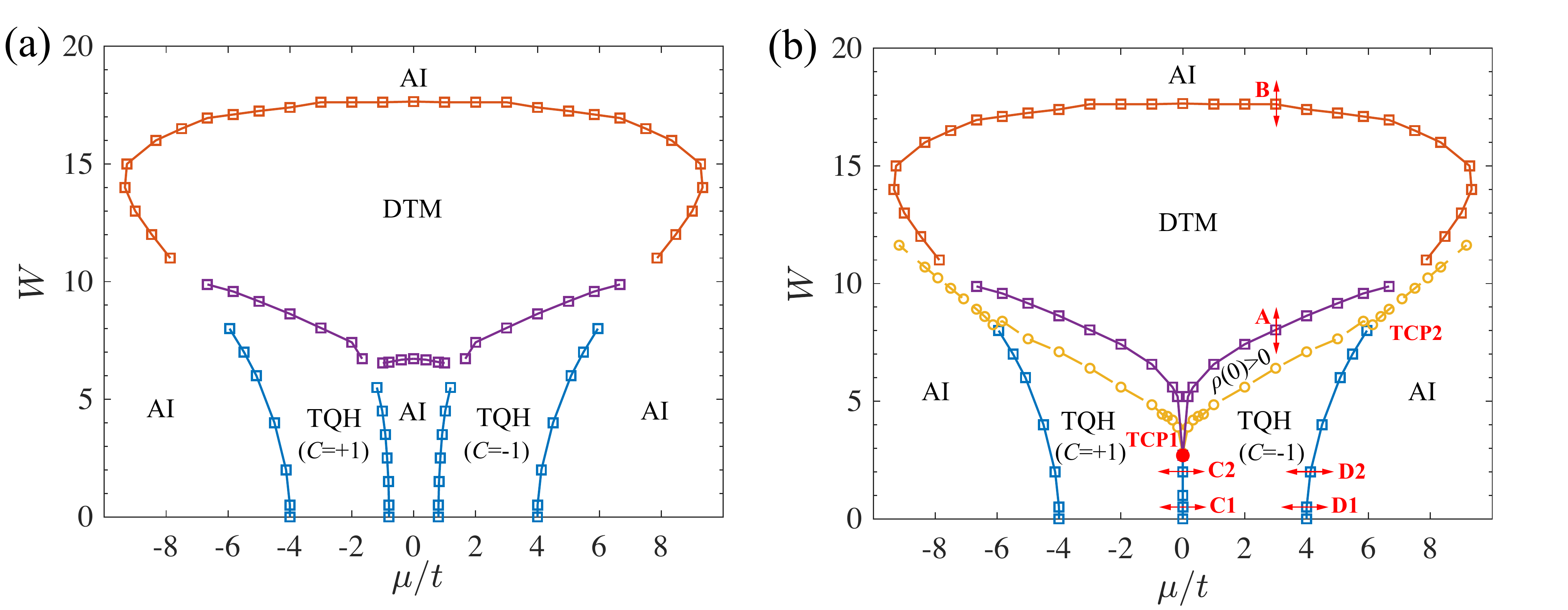}
\caption{Phase diagram of the class D disordered superconductor model, Eq.~(\ref{bdg}). (a) is for the anisotropic model at $\alpha=0.2$ and (b) is for the isotropic model ($\alpha=0$). The solid lines connecting squares are phase boundaries determined by the transfer matrix method. The yellow circles and dashed lines in (b) are determined by the kernel polynomial method with system size $L=3000$ and expansion order $ N=4000$. Zero-energy DOS takes non-negligible values ($\rho_{\rm KPM}(0)> 10^{-3}$) above this line, which we term a ``gap-closing''. TCP1 and TCP2 are tricritical points. The red arrows denote points and regions where we studied phase transitions: A and B are metal-insulator transitions at $\mu/t = 3$; C1, C2, D1, and D2 are thermal quantum Hall transitions at $W=0.5,\, 2,\, 0.5$ and $2$, respectively.}
\label{fig1}
\end{figure*}

The organization of the paper and its main findings are as follows.
In Section \ref{sec:phase-diagram}, we draw the phase diagram of the model Eq.~\eqref{bdg} (Fig.~\ref{fig1}). Panel (a) shows an anisotropic case with $\alpha = 0.2$, while panel (b) shows the isotropic case $\alpha = 0$. In both cases phase boundaries were identified by calculating the quasi-one dimensional (quasi-1D) localization length of zero-energy eigenstates using the transfer matrix method. The isotropic system has an enhanced $C_4$ lattice symmetry that causes two TQH transitions to merge into one on the vertical line $\mu = 0$. This symmetry makes the study of the isotropic system easier, and in the rest of the paper we focus on this case. We expect that universal properties of the phases and phase transitions are the same for $\alpha = 0$ and $\alpha \neq 0$.

In Section \ref{sec:DOS-metal}, we demonstrate a logarithmic divergence of the low-energy density of states (DOS) in the DTM phase using the kernel polynomial method \cite{KPM2006}. This result is consistent with theoretical predictions \cite{Senthil2000prb, Bocquet2000} and earlier numerical studies \cite{Mildenberger2007}. Our new result is that a logarithmic divergence of DOS is also present at the MIT point of the DTM-TQH boundary [point A in Fig.~\ref{fig1}(b)], which implies that the dynamical critical exponent $z$ at this transition is equal to the space dimension, $z=d=2$.

In Section \ref{sec:MI-transitions}, we evaluate the critical exponent of a divergent characteristic length  
$\nu$ at the MIT points of the DTM-TQH and DTM-AI boundaries (points A and B) using finite-size scaling analysis with polynomial fitting procedures~\cite{Slevin2014}. The values at the two points, $\nu_A = 1.35 \pm 0.04$ and $\nu_B = 1.36 \pm 0.05$, allow us to conclude that these transitions belong to the same universality class.

In Section \ref{sec:TQH-transitions}, we study thermal quantum Hall transitions between insulating phases with different quantized values of the thermal Hall conductivity
[points C1, C2, D1, and D2 in Fig.~\ref{fig1}(b)]. All these transitions exhibit very close values of the localization length exponents consistent with $\nu'=1$, the value at the clean Ising fixed point, confirming theoretical predictions \cite{Senthil2000prb, Chalker2001prb}.

In Section \ref{sec:TCP}, we study the critial properties at the vicinity of the tricritical point
where the phase boundary between distinct TQH phases at $\mu = 0$ terminates [point TCP1 in Fig.~\ref{fig1}(b)]. Even the determination of the position of the tricritical point in earlier numerical studies of network models was inconclusive \cite{Chalker2001prb, Mildenberger2007, Kagalovsky2008prl, Kagalovsky2010prb}. We establish the position of the tricritical point TCP1 from a scaling analysis of the longitudinal conductivity and the DOS along the critical line $\mu = 0$. Next, we determine a correlation length exponent $\nu''$ and the dynamical exponent $z''$ at TCP1. The exponents $\nu''$ and $z''$ at TCP1 turn out to be different from $\nu$ and $z$ at DTM-TQH and DTM-AI transitions (points A and B), indicating that TCP1 represents a distinct universality class. We conclude from these numerical observations that the TCP is an unstable fixed point with two relevant scaling variables in the $\mu-W$ plane.

The final section, Section \ref{sec:conclusions}, is devoted to summary and concluding remarks.

\section{The Phase Diagram}
\label{sec:phase-diagram}

Eq.~(\ref{bdg}) can also be expressed as
\begin{align}
	\mathcal{H}= \sum_{{\bm j},{\bm m}} (c_{\bm j}^{\dagger}  \!\ \!\ c_{\bm j})^{\vphantom{\dagger}} \mathbb{H}_{{\bm j},{\bm m}}  (c_{\bm m}^{\vphantom{\dagger}} \!\ \!\ c^{\dagger}_{\bm m})^T,
\label{bdg-matrix}
\end{align}
where the first-quantized Hamiltonian matrix $\mathbb{H}$ has the particle-hole symmetry
\begin{align}
	\sigma_1 \mathbb{H} \, \sigma_1 = - \mathbb{H}^T,
\label{PH-symmetry}
\end{align}
and the Pauli matrix $\sigma_1$ acts on the particle-hole space.

The model Eq.~\eqref{bdg-matrix} is easily solved in the clean limit ($\varepsilon_{\bm j} \equiv 0$) in the momentum space. Taking the lattice spacing to be 1, the two quasiparticle energy bands are
\begin{align}
E_{\pm}(k_x,k_y) = & \pm \big[(\mu + 2t_x \cos k_x + 2t_y \cos k_y)^2
\nonumber\\
& + 4\Delta^2 (\sin^2 k_x + \sin^2 k_y)\big]^{1/2}.
\label{band}
\end{align}
The spectrum is gapped except for $\mu = \pm 4t$ and $\mu = \pm 4\alpha t$. Quantization of the TKNN integer $C$ of the gapped quasiparticle bands
(BdG Chern number) results in the emergence of chiral Majorana edge modes, and a quantized thermal Hall conductance $\kappa_{xy} / T = C (\pi^2 k_B^2 /6h)$ \cite{Senthil2000prb}. For $|\mu|>4|t|$ and $|\mu|<4 |\alpha t|$, the system is an ordinary superconductor with $C=0$.
The regions $-4|t|<\mu<-4 |\alpha t|$ and $4 |\alpha t|<\mu<4|t|$ correspond to the two distinct topological superconducting phases (i.e., thermal quantum Hall phases) with $C=1$ and $-1$, respectively.

The spectrum Eq.~\eqref{band} contains four Dirac fermions at $(k_x, k_y) = (0,0)$, $(0,\pi)$, $(\pi, 0)$, and $(\pi,\pi)$. Quantum phase transitions between topologically distinct TQH phases at $\mu = \pm 4t$ and $\mu = \pm 4\alpha t$ correspond to the vanishing of the mass of one of the Dirac fermions. In the isotropic case as shown in Fig.~\ref{fig1}(b), the two topological transitions at $\mu=\pm 4 \alpha t$ merge into a single transition at $\mu=0$, where the BdG Chern number changes by 2. In this case the Dirac fermions that appear at $(k_x,k_y)=(\pi,0)$ and $(0, \pi)$ are related by the $C_4$ rotation symmetry.

\begin{figure}
\centering
\includegraphics[width=3.2in]{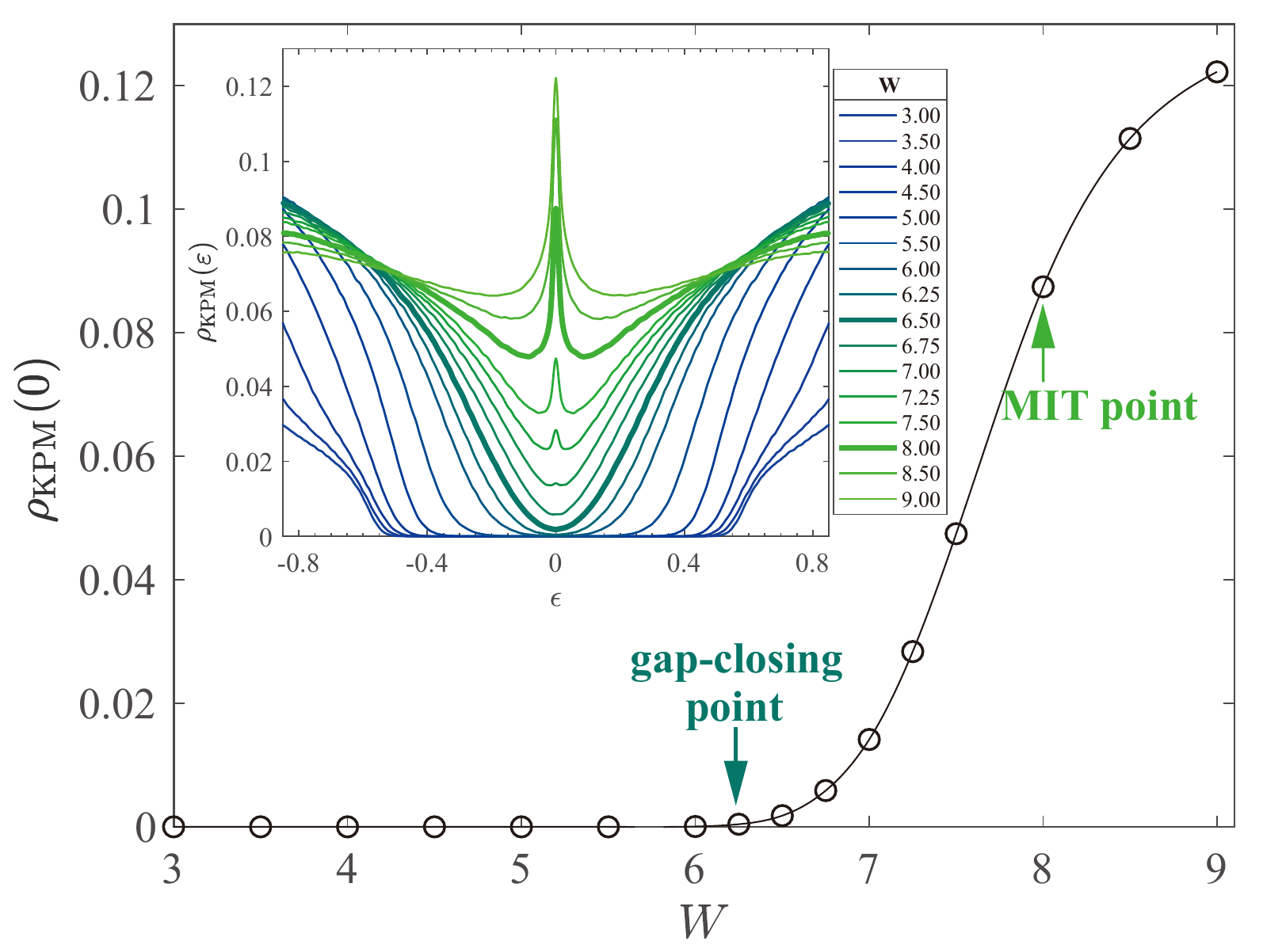}
\caption{Zero-energy DOS $\rho_{\rm KPM}(0)$ of the isotropic model as a function of disorder strength $W$ at $\mu/t=3$. The data points are calculated by the kernel polynomial method with $3000\times 3000$ lattice sites and expansion order $ N=4000$. Inset: $\rho_{\rm KPM}(\varepsilon)$ curves for different $W$ around $\varepsilon=0$. Disorder strength for the two bold arrows approximates the gap-closing point ($W_e \simeq 6.4$) and critical point for insulator-metal transition ($W_c = 8.03$), respectively.}
\label{fig2}
\end{figure}

To determine the phase diagram, we rewrite the model~\eqref{bdg} as a two-orbital model, see Appendix~\ref{appendix-A} for details. We then calculate the quasi-1D localization length $\lambda$ of zero-energy eigenstates by the transfer matrix method \cite{MacKinnon81, Pichard1981, MacKinnon83, Slevin2014}. A phase transition point is identified as a scale-invariant point of the normalized quasi-1D localization length $\Lambda \equiv \lambda/L$ with respect to various transverse size $L$, see Section \ref{sec:MI-transitions} for details. For numerical simulations, we use uniformly distributed disorder
\begin{align}
\varepsilon_{\bm j} &\in [-W/2,W/2],
& \overline{\varepsilon_{\bm{i}} \varepsilon_{\bm j}}
&= \delta_{\bm{i},\bm j}W^2/12,
\end{align}
and fix parameters $\Delta = 1$, $t = 0.6$ in Eq.~(\ref{bdg}).

The addition of week disorder ($W \lesssim 5$) localizes all bulk quasiparticle states in the ordinary and topological superconductors converting them to Anderson insulators (AIs) and thermal quantum Hall (TQH) phases, respectively. Direct transitions between distinct TQH phases remain intact, as shown in Fig.~\ref{fig1}. This observation is consistent with the one-loop renormalization group (RG) analysis of the 2D gapless Dirac fermion with random mass, where the random mass is a marginally-irrelevant perturbation at the clean Ising model fixed point \cite{Dotsenko1983}. In the strong disorder limit ($W \gtrsim 18$) the TQH phases disappear, and all the states are Anderson localized (AI phase). In the intermediate range of disorder strength ($8 \lesssim W \lesssim 17$) between the two extremes, there is a finite region of the DTM phase, where the quasiparticle eigenstates at zero energy are extended in the bulk. The range of disorder strength $W$ where the DTM phase exists is widest at $\mu/t = 0$, and gradually disappears when $|\mu|/t \gtrsim 9$.

The phase diagram of the isotropic model [Fig.~\ref{fig1}(b)] has similar structure to that of an anisotropic model [Fig.~\ref{fig1}(a)]. Note that the degeneracy of the two plateau transitions at $\mu=0$ in the isotropic case is not lifted by finite disorder, as the above-mentioned $C_4$ lattice symmetry is still present on average. The critical line along $\mu=0$ ends at a single tricritical point TCP1. The transfer matrix calculations of $\Lambda$ suffer from strong finite-size effect around tricritical points, making it difficult to analyze critical behavior in their vicinity (see Appendix \ref{appendix-B}). Nonetheless, we were able to determine the location and critical exponents at TCP1 using scaling analysis of the quasiparticle conductivity {\it along} the critical line.

\begin{figure}
\centering
\includegraphics[width=3.2in]{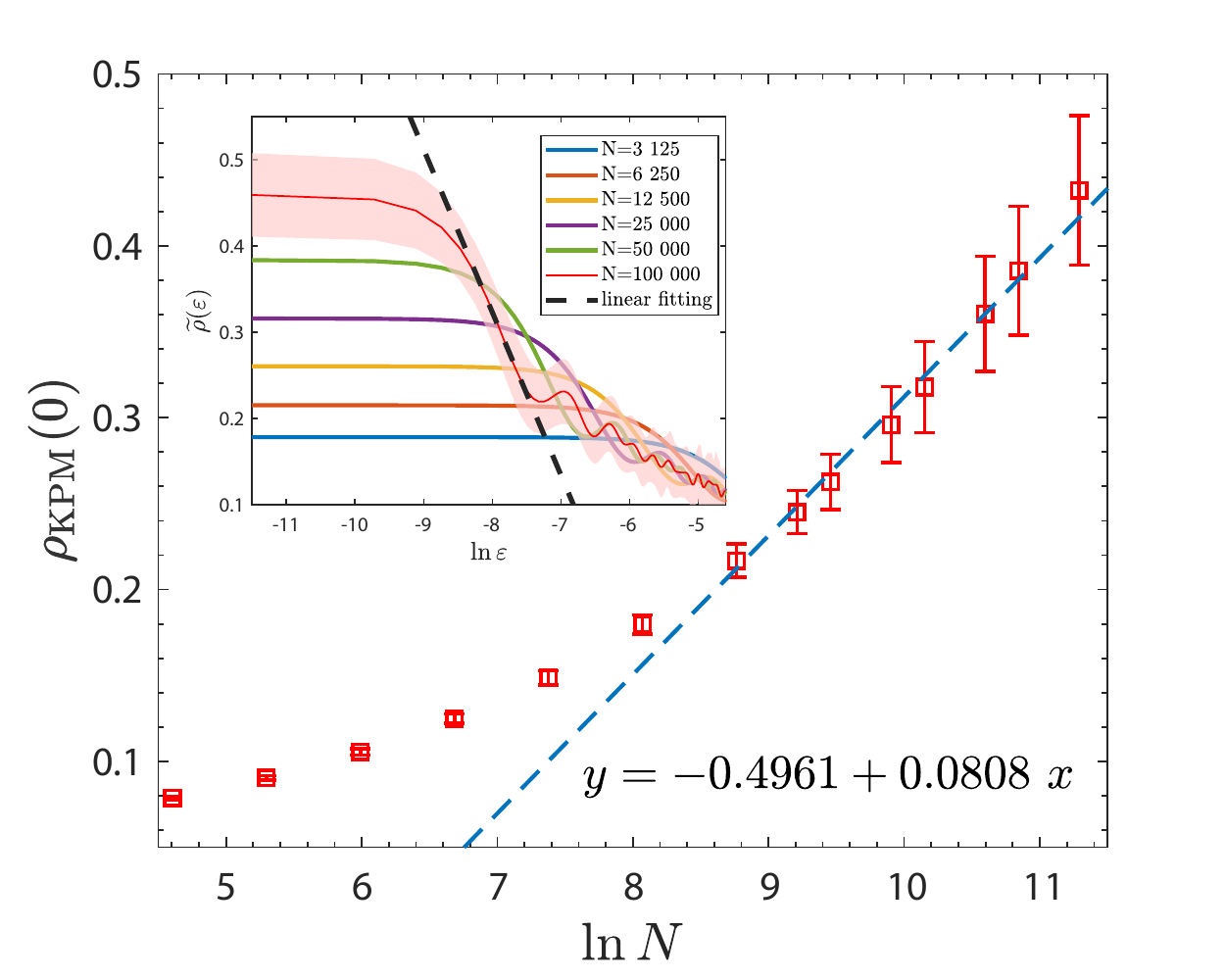}
\caption{Dependence of the zero-energy DOS on the kernel polynomial expansion order $N$ in the DTM phase. The data is taken at $(\mu/t,W)=(35/6,14)$ for the isotropic model of the square sample with $250\times 250$ lattice sites. The error bar is the standard deviation of 64 disorder configurations. The dashed line shows a linear fit to $\ln N$ in a region of  large $N$. Inset: $\rho_{\rm KPM}(\varepsilon)$ vs. $\ln \varepsilon$ for different values of $N$ at the same parameter point as in the main figure. The red shaded area is the error bar of the $N=100000$ curve, and the black dashed line is the linear fitting of data range $\varepsilon \in [3.5, 5.5] \times 10^{-5}$. }
\label{fig3}
\end{figure}

\section{Density of states in TQH and DTM phases and at DTM-TQH transition}
\label{sec:DOS-metal}

\begin{figure*}
\centering
\includegraphics[width=6.5in]{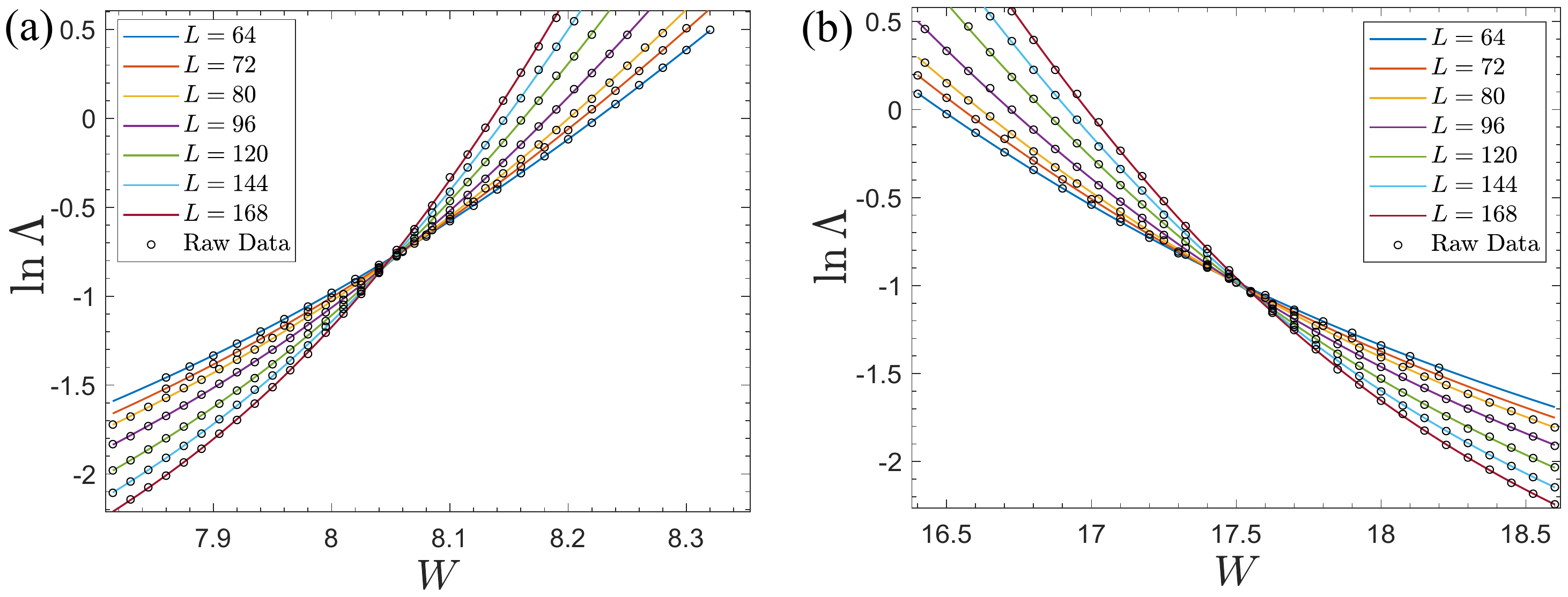}
\caption{Normalized quasi-1D localization length $\Lambda$ as a function of the disorder strength $W$ at $\mu/t=3$ for different transverse size $L$.
(a): near TQH-DTM transition point [point A in Fig.~\ref{fig1}(b)].
(b): near the DTM-AI transition point [point B in Fig.~\ref{fig1}(b)]. Black circles are raw data points, and colored curves are the polynomial fitting curves with the largest goodness of fit. The both fitting curves for (a) and (b) are obtained with the expansion order of $(n_1, n_2, m_1, m_2) = (3, 1, 2, 0)$.}
\label{fig4}
\end{figure*}

The density of states provides complementary information about the phases and phase transitions. For simplicity we focus on the isotropic model only, and impose periodic boundary conditions to get the density of the bulk states. Calculations by the kernel polynomial method show that the zero-energy DOS in TQH phases is vanishingly small at weak disorder, but acquires a finite value when the disorder strength exceeds a certain value $W_e$, which we call the gap-closing point or simply gap-closing. Increasing disorder strength beyond $W_e$, we observe a transition to DTM phase at a critical value $W_c$, which corresponds to the DTM-TQH transition point. For $\mu/t = 3$ the gap-closing point is at $W_e \simeq 6.4$, and the DTM-TQH transition is at $W_c = 8.03$, as shown in Fig.~\ref{fig2}. Details of the precise determination of $W_c$ are explained in the following section, Sec.~\ref{sec:MI-transitions}.

Determining $W_e$ and $W_c$ for many values of $\mu/t$, we observe that the gap-closing and the MIT become closer in the vicinity of tricritical points, and possibly merge at these points. A similar separation and merging of the gap-closing and MIT were observed around a tricritical point in a model of a three-dimensional (3D) disordered semimetal \cite{Luo2018prb}.

In the DTM phase, one can study the DOS and other properties using a non-linear sigma model approach \cite{Senthil2000prb, Bocquet2000, Mildenberger2007, Evers2008RMP}. The study results in a logarithmic divergent DOS around the zero single-particle energy ($\varepsilon=0$),
\begin{align}
   \rho(\varepsilon) \propto \ln
   \frac{1}{|\varepsilon|}
   + {\cal O}(1).   \label{rhoE}
\end{align}
The logarithmic divergence is numerically confirmed in Fig.~\ref{fig3}. The inset shows the DOS around $\varepsilon=0$ calculated with different values of the kernel polynomial expansion order $N$. In the kernel polynomial calculation~\cite{KPM2006} of the DOS $\rho(\varepsilon) \equiv \frac{1}{V} \sum_{i} \delta(\varepsilon-\varepsilon_i)$, the $\delta$-function is expanded in terms of Chebyshev polynomials up to the order $N$. Thereby, the $\delta$-function is approximated by
\begin{align}
\delta_N(\varepsilon - \varepsilon_i) = \frac{a}{\pi N} 
\frac{1}{(\varepsilon-\varepsilon_i)^2 + (a N^{-1})^2}
\end{align}
for finite large $N$, and the energy resolution is limited by $aN^{-1}$. $a$ is a 
coefficient of order unity. 
Therefore, $\rho_{\rm{KPM}}(\varepsilon)$ increases logarithmically with $|\varepsilon|^{-1}$ only for $|\varepsilon|\gg N^{-1}$, while it converges to a constant value $\rho_{\rm{KPM}}(0)$ for $|\varepsilon|\ll N^{-1}$. When $N$ increases, the constant value $\rho_{\rm{KPM}}(0)$ increases logarithmically,
\begin{align}
\rho_{\rm KPM}(N,\varepsilon=0)\propto \ln (N) + {\cal O}(1).
\label{rhoN}
\end{align}
Details can be found in Appendix \ref{appendix-D}.

The main part of Fig.~\ref{fig3} demonstrates that the zero-energy DOS in the DTM phase indeed increases linearly with $\ln N$. Previous studies of the non-linear sigma model also predicted logarithmic divergence of conductance in the limit of large system size~\cite{Evers2008RMP}. Numerical results on the Landauer conductance in the DTM phase are consistent with this prediction (see Appendix \ref{appendix-C}).

Fig.~\ref{fig2} suggests that the DOS at the DTM-TQH metal-insulator transition (MIT) point, as well as in the localized phases nearby, also have weak divergences around zero energy. Calculating $\rho_\mathrm{KPM}(0)$ with different $N$, we confirmed that the zero-energy DOS at the DTM-TQH MIT point
scales as $\ln N$ for large $N$, while it scales as $N^{-\alpha}$ with a non-universal exponent $\alpha$ in the localized phases near the MIT point (see Appendix \ref{appendix-D}). The divergence of $\rho_\mathrm{KPM}(0)$ as a function of $N$ implies the same kind of divergence of $\rho(\varepsilon)$ as a function of $\varepsilon$.

The logarithmic scaling of the low-energy DOS at the MIT point means that the dynamical critical exponent is the same as the space dimension: $z=d=2$. In that regard, the type of MIT of the DTM-TQH boundary in the class D model is not different from the Anderson transitions in the standard
Wigner-Dyson symmetry classes, where $z=d$. The power-law scaling of the low-energy DOS in the localized phases could be related to Griffiths singularities, as suggested by a similar power-law divergence found in a 2D class-D network model~\cite{Mildenberger-Griffiths-2006}.

\section{Scaling behavior at metal-insulator transitions}
\label{sec:MI-transitions}

\begin{table*}
\caption{Results of finite-size scaling analysis at the metal-insulator transitions A and B along the line of $\mu/t=3$. The Taylor expansion orders in Eqs. (\ref{eq:u1u2}) and (\ref{eq:Lambda}) are chosen as $(n_1, n_2, m_2) = (3, 1, 0)$, while $m_1$ is either 2 or 3. The fittings are also carried for different ranges of transverse size $L$. The square brackets denote 95\% confidence intervals from 1000 Monte Carlo simulations.}
	\flushleft
	(a) transition A (TQH-DTM transition)
	\begin{ruledtabular}
		\begin{tabular}{cccccccc}
			$m_1$ & $L$ & $W$ & GOF & $ W_c$ & $\nu$ & $y$ & $\Lambda_c$ \\
			\hline
			2 & $\ge 56$ & [7.8, 8.32] & 0.76 & 8.026[8.016, 8.034] & 1.371[1.311, 1.437] & 0.789[0.443, 1.266] & 0.346[0.293, 0.389] \\
			2 & $\ge 64$ & [7.8, 8.32] & 0.90 & 8.028[8.016, 8.036] & 1.351[1.262, 1.408] & 0.910[0.462, 1.584] & 0.358[0.305, 0.401] \\
			3 & $\ge 56$ & [7.8, 8.32] & 0.85 &8.024[8.016, 8.032] & 1.363[1.287, 1.434] & 0.696[0.450, 1.112] & 0.333[0.292, 0.379] \\
			3 & $\ge 64$ & [7.8, 8.32] & 0.89 & 8.026[8.014, 8.035] & 1.342[1.231, 1.411] & 0.784[0.430, 1.541] & 0.344[0.302, 0.396] \\
		\end{tabular}
	\end{ruledtabular}
	\flushleft
	(b) transition B (DTM-AI transition)
	\begin{ruledtabular}
		\begin{tabular}{cccccccc}
			$m_1$ & $L$ & $W$ & GOF & $ W_c$ & $\nu$ & $y$ & $\Lambda_c$ \\
			\hline
			2 & $\ge 56$ & [16.6, 18.6] & 0.34 & 17.612[17.582, 17.647] & 1.348[1.279, 1.402] & 1.062[0.593, 1.747] & 0.311[0.282, 0.333] \\
			2 & $\ge 64$ & [16.6, 18.6] & 0.46 & 17.624[17.585, 17.694] & 1.360[1.241, 1.448] & 1.013[0.317, 2.080] & 0.303[0.256, 0.333] \\
			3 & $\ge 56$ & [16.4, 18.6] & 0.43 & 17.612[17.581, 17.639] & 1.332[1.274, 1.393] & 1.088[0.676, 1.887] & 0.311[0.282, 0.334] \\
			3 & $\ge 64$ & [16.4, 18.6] & 0.48 & 17.623[17.584, 17.663] & 1.360[1.286, 1.457] & 1.047[0.556, 2.118] & 0.304[0.261, 0.333] \\
		\end{tabular}
	\end{ruledtabular}
\label{tab1}
\end{table*}

The DTM-TQH and the DTM-AI transitions are both Anderson-type metal-insulator transitions characterized by the power-law divergence of the characteristic length 
$\xi \sim |x-x_c|^{-\nu}$. 
Here $x$ is a tuning parameter with a critical value $x_c$ at the Anderson transition point. In this section, we choose $x$ to be the disorder strength $W$. We evaluate the critical exponent $\nu$ for both DTM-TQH and DTM-AI transitions by a finite-size scaling analysis and polynomial fitting procedure of a normalized quasi-1D localization length $\Lambda$~\cite{Slevin1999correction, Slevin2014}. A standard argument leads to the scaling form $\Lambda=F(\phi_1,\phi_2)$, where
\begin{align}
\phi_1 &\equiv u_1(w) L^{1/\nu},
&
\phi_2 &= u_2(w) L^{-y},
\end{align}
are the relevant and the least irrelevant ($y > 0$) scaling variables near the fixed point controlling the transition. 
For the  
disorder-induced transition, $u_{1,2}$ are functions of the dimensionless disorder $w = (W-W_c)/W_c$.

\begin{figure}
\centering
\includegraphics[width=3.2in]{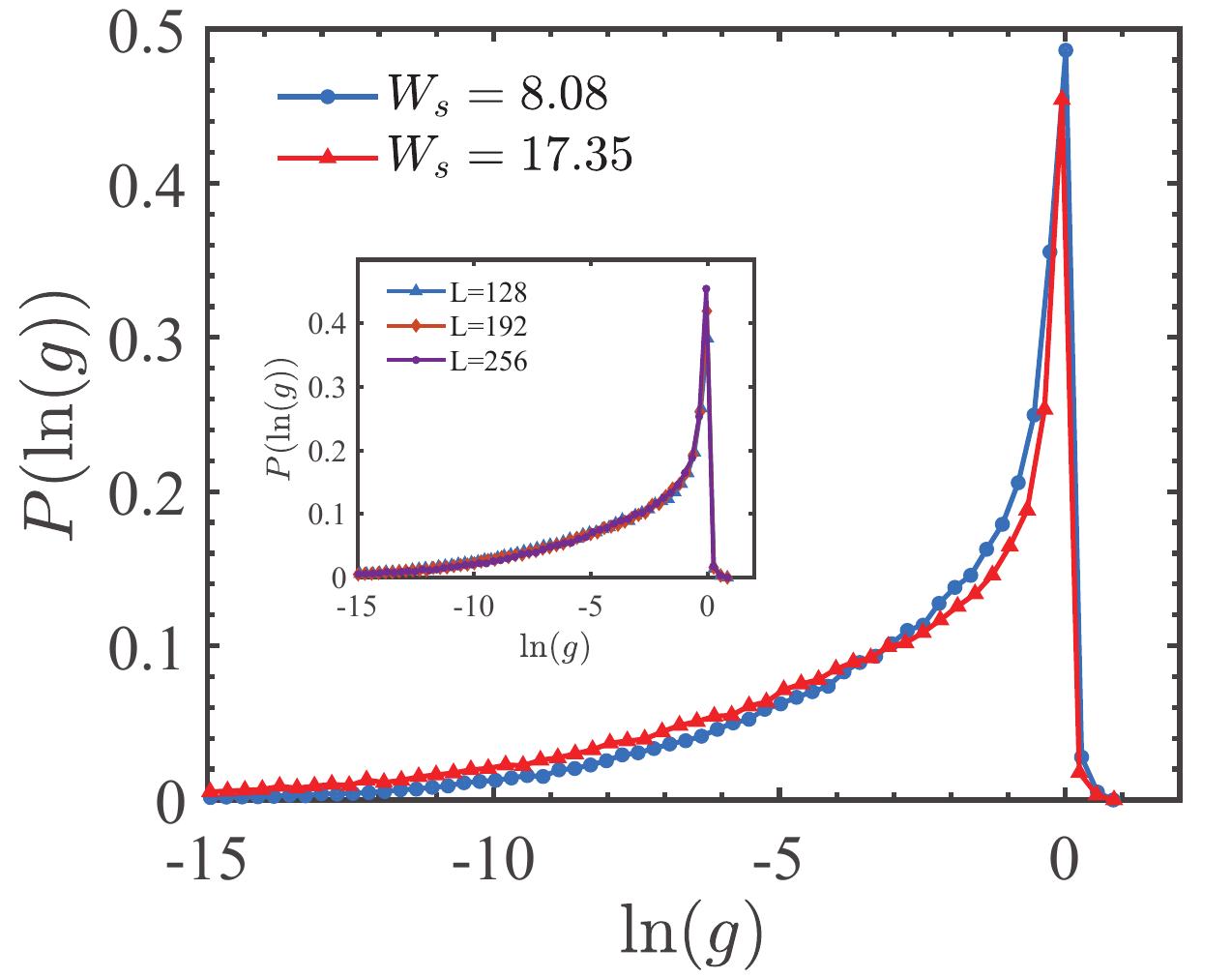}
\caption{Distributions of the two-terminal conductance at the DTM-TQH transition point (blue curves) and at the DTM-AI transition point (red curves). The critical conductances are computed in the square geometry ($L\times L$,
$L=128,192,256$) with the periodic boundary condition along the transverse direction. The distributions are calculated at $W_s = 8.08$ (for the DTM-TQH transition) and $W_s=17.35$ (for the DTM-AI transition) on the line $\mu/t=3$. We chose the values $W_s$ for the two transitions so that $\langle g\rangle$ is scale-invariant. This is illustrated in the inset, where the critical conductance distributions calculated with three different system sizes are seen to overlap well at $W_s=8.08$. Note that $W_s$ thus determined are slightly different from the corresponding critical disorder strengths $W_c$ determined by the polynomial fitting.}
\label{fig5}
\end{figure}

\begin{figure*}
\centering
\includegraphics[width=6.5in]{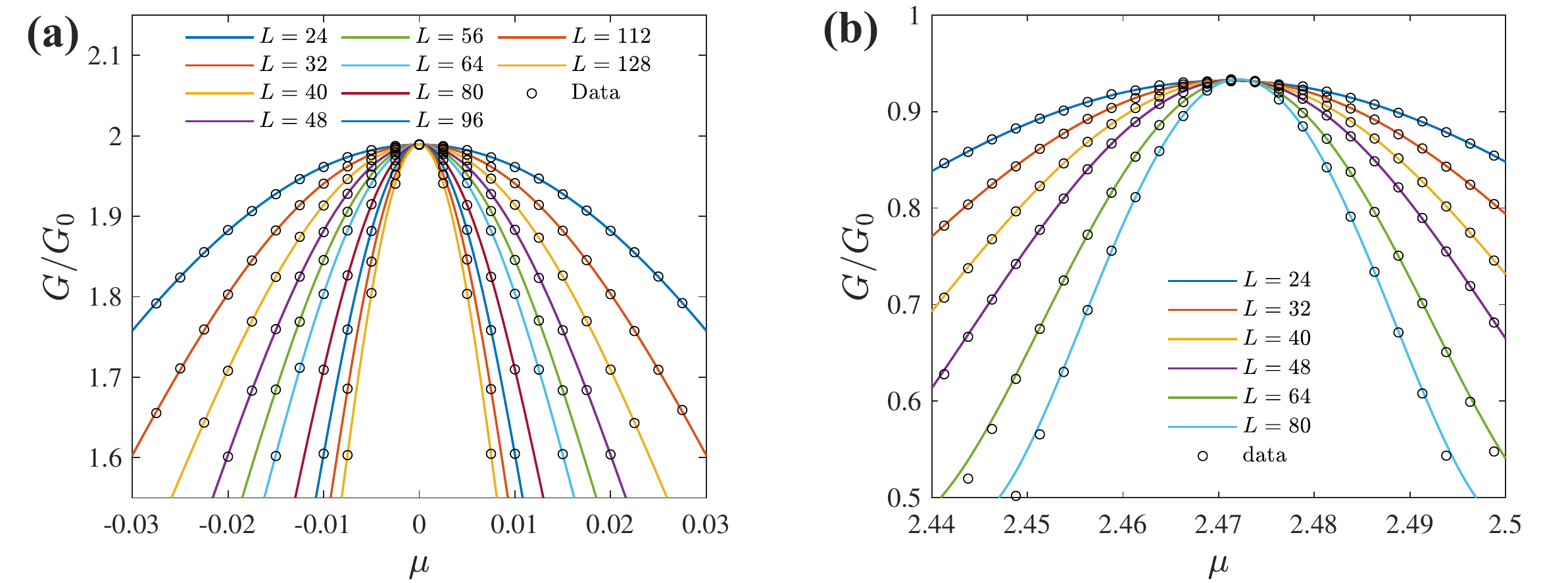}
\caption{Two-terminal conductance $G$ as a function of $\mu$ near (a) transition C1 ($W=0.5$) and (b) transition D2 ($W=2$). $G_0=\pi^2 k_B^2 T/6h$ is the thermal conductance quantum. The black circles are raw data points and the colored curves are from the fitting results. Each data point is averaged over a few tens of thousands samples to guarantee the precision of 0.1\%.}
\label{fig6}
\end{figure*}

In the multiple-dimensional parameter space of parameters (which include $w$), the equation $w = 0$ defines the critical subspace where $\phi_1 = 0$. On this subspace there is a fixed point controlling the Anderson transition, where all irrelevant scaling variables, including $\phi_2$, vanish. Near the critical subspace both $u_1(w)$ and $u_2(w)$ can be expanded in Taylor series in powers of $w$ :
\begin{align}
u_1(w) &= \sum_{j=1}^{m_1} b_j w^j,
&
u_2(w) &= \sum_{j=0}^{m_2} c_j w^j.
\label{eq:u1u2}
\end{align}
For sufficiently large $L$ and small $w$, both $\phi_1$ and $\phi_2$ are
small and the scaling function can be further expanded near the fixed point as
\begin{align}
F = \sum_{j=0}^{n_1}\sum_{k=0}^{n_2}a_{j,k} \phi_1^j \phi_2^k\,
\label{eq:Lambda}
\end{align}
with $a_{1,0}=a_{0,1}=1$.
For a given set of expansion orders $(n_1,n_2,m_1,m_2)$, we minimize
\begin{align}
\chi^2 \equiv \sum_{n=1}^{N_D} \frac{(F_n - \Lambda_n)^2}{\sigma_n^2}
\end{align}
using $W_c, \nu, y, \{a_{j,k}\}, \{b_j\}$ and $\{c_j\}$ as fitting parameters. Here $N_D$ is the number of data points, $\Lambda_n$ and $\sigma_n$ are the $n$-th data point and its standard error, and $F_n$ is the fitting value from
Eq.~(\ref{eq:Lambda}).
We perform the fittings with $(n_1,n_2,m_2)=(3,1,0)$ and $m_1=2,\!\ 3$,
that give a goodness of fit (GOF) well over 0.1. Results of such fittings for the two metal-insulator transitions are shown in Fig.~\ref{fig4}
and Table \ref{tab1}.

As shown in Table \ref{tab1}, the fitting results are stable against changes in $m_1$ and the range of system sizes $L$. The results are
\begin{align}
\nu &= 1.35 \pm 0.04 & &\text{for DTM-TQH},
\\
\nu &= 1.36 \pm 0.05 & &\text{for DTM-AI}.
\end{align}
Each value is taken from the fitting with the largest GOF, and the numbers after the $\pm$ signs are the standard deviations estimated by Monte Carlo simulations. Our value for $\nu$ at DTM-TQH transition agrees with that of Ref. \cite{Kagalovsky2008prl}, which reported $\nu = 1.4 \pm 0.2$. The 95\% confidence intervals of $\nu$ at the two transitions overlap within rather small errors bars, suggesting that the transitions belong to the same universality class.

To reinforce this conclusion, we further compare the distributions of the quasiparticle (thermal) conductance at the two transitions. The conductance distribution should be scale-invariant at an Anderson metal-insulator transition point \cite{Shapiro1990CCD}, and should only depend on the universality class and the sample geometry \cite{Slevin1997TRS}. Using the transfer matrix method \cite{Pendry1992, Kramer2005rev}, we calculated the two-terminal conductance $g$ of $10^6$ samples with square geometry and the periodic boundary condition in the transverse direction. Figure \ref{fig5} shows the critical conductance distributions at the DTM-TQH and the DTM-AI transitions. The distributions match well with each other. The critical conductance
distributions and the localization length exponents $\nu$ at the two transitions strongly suggest
that the two metal-insulator transitions are in the same universality class. We note that this is similar to the case of the quantum spin Hall (QSH) systems, where it is known that the critical behaviors at the transition between the diffusive metal (DM) and the QSH phase is the same as those at the DM-AI transition~\cite{Obuse07, Kobayashi10, Fu12}.

\begin{table*}
\caption{Results of a finite-size scaling analysis of the two-terminal conductance at TQH transitions C1, C2, D1 and D2. The conductance is calculated with the square geometry ($L\times L$) and
with the periodic boundary condition along the transverse direction. For C1, C2 and D1, we omit the dependence
of the conductance scaling function on the irrelevant scaling variable, therefore $n_2 = 0$ in Eq.~(\ref{eq:Lambda}). The expansion orders associated with the relevant scaling variable are fixed to $(n_1,m_1) = (4,3)$.
The square brackets denote 95\% confidence intervals evaluated from 1000 Monte Carlo simulations.
$\mu_c = 0$ for transitions C1 and C2 by the symmetry (see the text).}
	\begin{ruledtabular}
		\begin{tabular}{cccccccccc}
			\ & $L$&$G$&$n_2$&$m_2$&GOF& $\mu_c$ & $\nu'$ & $y'$ & $G_c/G_0$  \\
			\hline
			C1&24$-$128& $>$1.6 &0&-&0.14& 0 &1.002\,[0.999,\,1.003]& - &1.9891\,[1.9890,\,1.9892]\\
			C2&24$-$80 & $>$1.45 &0&-&0.11& 0 &1.017~[1.009,1.026]& - &1.8494\,[1.8490,\,1.8498]\\
			D1&24$-$128& $>$0.8 &0&-&0.34&2.4045\,[2.4045,\,2.4046]&1.000\,[0.998,\,1.002]& - &0.9947\,[0.9946,\,0.9947]\\
			D2&24$-$80 & $>$0.7 &1&0&0.13&2.4727\,[2.4726,\,2.4729]&1.008\,[0.964,\,1.020]& 0.59\,[0.51,0.68]&0.9343\,[0.9334,\,0.9351]\\
		\end{tabular}
	\end{ruledtabular}
\label{tab2}
\end{table*}

\section{Thermal quantum Hall transitions}
\label{sec:TQH-transitions}

The TQH-TQH transition and TQH-AI transition are both direct insulator-insulator transitions (plateau transitions) characterized by the power-law divergence of the localization length $\xi \sim |x - x_c|^{-\nu^{\prime}}$. In this section we take $\mu/t$ as the variable $x$. When varying $\mu/t$ with fixed weak disorder $W \lesssim 5$, we encounter a sequence of thermal quantum Hall transitions between topologically distinct TQH phases in the phase diagram in Fig.~\ref{fig1}. Critical properties at these plateau transitions have been reported in previous literature \cite{Kagalovsky2008prl, Kagalovsky2010prb, Medvedyeva2010prb}, but the numerical results are not consistent. For the isotropic model Eq.~(\ref{bdg}), it is also not obvious whether the TQH-TQH transition at $\mu=0$ has the same universal properties as the TQH-AI transition around $\mu=\pm 4t$, because the former transition is described by a field theory of two copies of disordered Dirac fermions, as opposed to one copy of Dirac fermions for the latter transition.

To compare the scaling behavior at the two TQH transitions in the isotropic model, we employ a finite-size scaling analysis for the two-terminal conductance, and evaluate the localization-length exponent $\nu'$ from its polynomial fitting. At TQH transition points, both the normalized quasi-1D localization length $\Lambda$ and the two-terminal conductance should show a scale-invariant behavior. However, $\Lambda$ diverges at TQH transitions in the clean limit. In the weak disorder limit, this divergence is cut off 
by finite system sizes, 
but the critical value of $\Lambda$ is still large, and also exhibits large statistical errors. At the same time, the average critical conductance goes to a constant at a TQH transition, with smaller statistical errors (see Appendix \ref{appendix-B}). We thus use the dimensionless average conductance $G = \langle g \rangle$ as the scaling quantity for the evaluation of $\nu^{\prime}$.

The conductance $g$ is calculated in a square geometry, $L\times L$, with the periodic boundary condition in the transverse direction. For each $\mu$ and $L$, we take an average over at least $10^4$ samples, to guarantee $0.1\%$ precision for the average conductance $G = \langle g \rangle$. Near the TQH transition point, $G$ can be fitted by a scaling function of the relevant scaling variable $\phi_1$ and the least irrelevant scaling variable $\phi_2$ as
$G = F(\phi_1, \phi_2)$. For fixed $W$, these scaling variables are
\begin{align}
\phi_1 &= u_1(\delta\mu) L^{1/\nu'},
&
\phi_2 &= u_2(\delta\mu)L^{-y'},
\end{align}
where $\delta \mu \equiv \mu-\mu_c$ is the deviation of $\mu$ from its critical value $\mu_c$ at a TQH transition. In the vicinity of the fixed point (where $\phi_1 = \phi_2 = 0$), both $u_1$ and $u_2$ can be expanded in small $\delta \mu$. We introduce $m_1$ and $m_2$ as the respective Taylor expansion orders
as in Eq.~(\ref{eq:u1u2}).

We choose two representative critical points at each of the two TQH
critical lines in the isotropic model, one at weak disorder [points C1, D1 in Fig.~\ref{fig1}(b)] and one at a stronger disorder [points C2, D2 in Fig.~\ref{fig1}(b)]. For the transitions C1, C2 and D1, corrections due to
irrelevant scaling variables are negligible [Fig.~\ref{fig6}(a)],
and the data can be well fitted by a single-parameter scaling function $G = F(\phi_1)$, i.e. $n_2=0$. For the transition D2, both the relevant and the least irrelevant scaling variables are essential due to a considerable
finite-size corrections [Fig.~\ref{fig6}(b)].

For the transitions C1 and C2, the scaling function $F(\phi_1)$ is even in $\phi_1$ because the conductance $G$ is an even function of $\mu$ due to the particle-hole symmetry. Indeed, the first-quantized Hamiltonian matrix is Hermitian, and the particle-hole symmetry Eq.~\eqref{PH-symmetry} can be rewritten as $\mathbb{H}^{*} = - \sigma_1 \mathbb{H} \sigma_1$. If we write $\mathbb{H}$ explicitly as a function of $\mu$, $\{\varepsilon_{\bm j}\}$, $t_{x}$, $t_y$, and $\Delta$, the symmetry becomes:
\begin{align}
    \mathbb{H}^{*}(\mu,\{\varepsilon_{\bm j}\},t_{\nu},\Delta)
    = \sigma_1 \mathbb{H}(-\mu,\{-\varepsilon_{\bm j}\},-t_{\nu},-\Delta) \sigma_1.
    \label{time-reversal}
\end{align}
The signs of $t_{x}$, $t_y$ and $\Delta$ can be further reversed by a gauge transformation that assigns $-1$ for one of the two sublattices in the square lattice. Therefore, if $\{ -\varepsilon_{\bm j}\}$ and $\{\varepsilon_{\bm j}\}$ appear with equal probabilities in the ensemble of different disorder realizations, all physical quantities that are even under the complex conjugation, including the average conductance, should be even functions of $\mu$. Thus, for the transitions C1 and C2 that happen at $\mu_c = 0$, we keep only even terms in the expansion of Eq.~(\ref{eq:Lambda}) in $\phi_1$, and only odd terms in the expansion of Eq.~(\ref{eq:u1u2}) in $\delta \mu$ as in the quantum Hall transition\cite{Slevin09}.
For the
transitions D1 and D2, we keep both even and odd terms in $\phi_1$ and $\delta \mu$. The results of the best fits for these four plateau transitions are summarized in Table.~\ref{tab2}.

The localization-length exponents $\nu'$ at all four transitions are very close to $\nu' = 1$, consistent with the theoretical prediction that a TQH transition is controlled by the clean Ising fixed point, where the disorder is marginally
irrelevant~\cite{Senthil2000prb, Chalker2001prb}. This also suggests that any mixing between the two copies of Dirac fermions at $\mu = 0$ due to disorder does not change the critical nature of the TQH plateau transition, compared to the one copy of Dirac fermion at $\mu/t=4$. Therefore, we expect that we are able to study the scaling properties near tricritical points of 2D class D disordered superconductors by focusing only on TCP1 in the next section.

\begin{figure}
\centering
\includegraphics[width=3.2in]{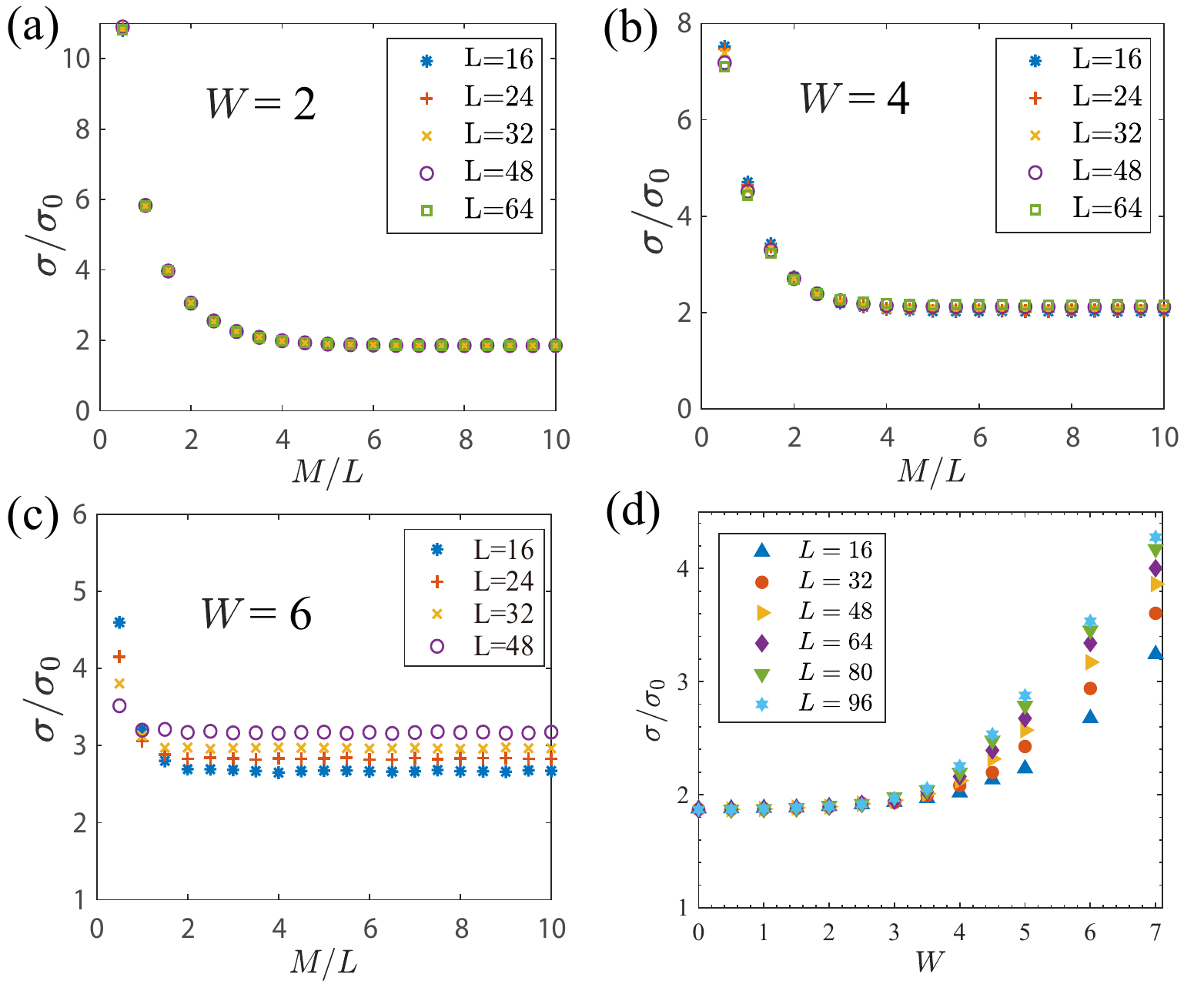}
\caption{(a)-(c) Thermal conductivity $\sigma \equiv GL/M$
as a function of sample aspect ratio $R\equiv M/L$. The thermal conductance
$G$ is calculated in a cylinder geometry of length $L$ and
circumference $M$ in the isotropic model along $\mu=0$
and different disorder strength $W=2,4$, and $6$, respectively
($\sigma_0 = G_0/\pi$).
Each data point is an average over $\sim 10^2 - 10^3$ disorder
configurations to reach the precision 0.2\%.
(d) $\sigma$ at fixed (sufficient large) aspect ratio $R=5$
as a function of $W$ along the $\mu=0$ line for different length
$L$.}
\label{fig7}
\end{figure}

\begin{figure}[t]
\centering
\includegraphics[width=3.in]{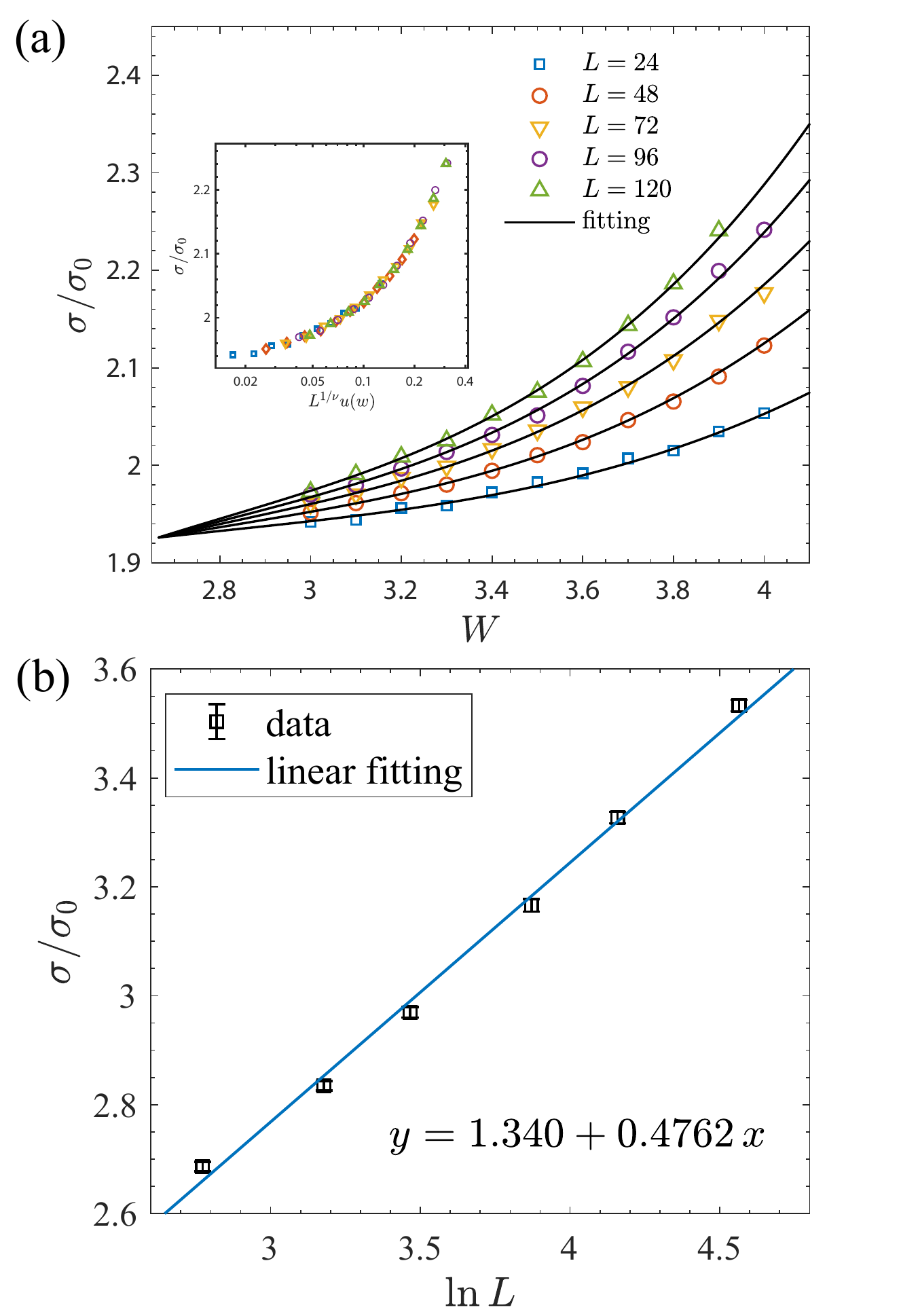}
\caption{(a) Conductivity $\sigma$ as a function of $W$ along the $\mu=0$ line, with a fixed aspect ratio
$M/L=5$. Each data point (colored markers) is the average over a few thousand disorder realizations. The black lines are from the polynomial fitting, where the data in a disorder range of $3<W<4$ are fitted by Eq.~(\ref{eq:sigma_fitting}) with the expansion order $(n, m)=(1,3)$. Inset: single-parameter scaling function for $\sigma=f(L^{1/\nu''}u(w))$, where we used $\nu^{\prime\prime}=1.54$ and $W_c=2.67$.
All the data in the range of $3<W<4$ and different length $L$ collapse into the single scaling function.
(b) Conductivity as a function of $\ln L$ at $(\mu/t,W)=(0,6)$. The
conductivity is obtained from the two-terminal conductance $G$ of the cylinder geometry with a length $L$ and circumference $M$, $\sigma = GL/M$, with fixed aspect ratio $M/L=3$. The error bar is the standard error of $10^2-10^3$ samples. The relative precision is around $0.2\%$. 
}
\label{fig8}
\end{figure}

\section{Tricritical point TCP1}
\label{sec:TCP}

The MIT lines of the DTM-TQH and DTM-AI boundaries, and the plateau transition lines for the AI-TQH and TQH-TQH boundaries merge into tricritical points in the phase diagrams. TCP1 and TCP2 are such multicritical points in the isotropic model [Fig.~\ref{fig1}(b)], where the DTM, TQH with $C=+1$ and TQH with $C=-1$ phases share their boundaries at TCP1, and the DTM, TQH with $C=1$ and AI with $C=0$ phases share their boundaries at TCP2. As we mentioned in the introduction, the critical behavior at the tricritical points in 2D class D systems is not well understood. Here, using the isotropic model, we locate the position of TCP1 on the phase boundary at $\mu = 0$ with high precision, and study the critical behavior at this point in detail. The idea of our approach is the following.

The plateau transition between topologically distinct clean superconductors (i.e., TQH conductors) at $\mu = 0$ is stable against weak disorder, since weakly-random mass is renormalized to zero for Dirac fermions. Thus, the vertical critical line $\mu = 0$ with sufficiently small $W$ represents the Dirac semimetal phase. Quasiparticle transport on this line is ballistic, and the (thermal) conductivity $\sigma$ should show a scale-invariant quantized value determined by the number of Dirac fermions \cite{Tworzydlo2006, Katsnelson2006, Schuessler2009prb}. On the other hand, beyond a certain critical value $W_\text{TCP}$, the system enters the DTM phase where the conductivity should become logarithmically divergent with the system size \cite{Senthil2000prb}. Based on this picture, we study the conductivity $\sigma$ as a function of the system size $L$ along $\mu = 0$ line and determine $W_\text{TCP}$ as the point where $\sigma$ starts to show a significant $L$ dependence.

We calculate the thermal conductance $G$ of cylindrical systems of length $L$ and circumference $M$ in the isotropic model along $\mu=0$ by the transfer matrix method. We then convert the conductance to conductivity, using
$\sigma \equiv G L/M$. Fig.~\ref{fig7}(a)-(c) show $\sigma$ as a function of
the aspect ratio $M/L$ for several values of the disorder $W$ and system lengths $L$. We see that for each $W$ and $L$, the conductivity data approach a constant as $M/L$ is increased, and that $M/L = 5$ is already large enough to approximate the $M/L \to \infty$ limit.

Fig.~\ref{fig7}(d) shows $\sigma$ at a fixed aspect ratio $M/L = 5$ as a function of the disorder strength $W$ for different system sizes $L$.
The conductivity barely changes with $L$ as long as the disorder strength is
below a certain critical value $W_\text{TCP}$. The comparison of the conductivity at $\mu = 0$ and $\mu = 4t$ in the clean limit confirms the quantization of $\sigma$ to the number of Dirac nodes (data not shown). Above $W_\text{TCP}$ the conductivity increases significantly with the system size $L$. These observations suggest that the portion $W > W_\text{TCP}$ on the $\mu=0$ line is already in the DTM phase, and $(W,\mu) = (W_\text{TCP},0)$ corresponds to the disorder-induced semimetal-metal quantum phase transition tricritical point [TCP1 in Fig.~ \ref{fig1}(b)].

This phase transition is characterized by the power-law divergence of a characteristic length scale $\xi$ on the DTM side ($W>W_{\rm TCP}$), $\xi \sim (W-W_{\rm TCP})^{-\nu''}$. To precisely determine $W_\text{TCP}$ and the exponent $\nu''$ at TCP1, we use a one-parameter scaling function for the conductivity in the DTM phase, $\sigma(L,W) = f(L^{1/\nu''}u)$, where
\begin{align}
u(w) &= \sum_{k=1}^{m} b_k w^{k},
&
w &= \frac{W-W_\text{TCP}}{W_\text{TCP}}
\label{u-w}
\end{align}
is the relevant scaling variable in the $\mu=0$ subspace.
Note that this scaling form is valid only on the DTM side ($w > 0$). Near TCP1, we expand $f(x)$ in powers of small $x$:
\begin{align}
	\sigma(L,w) = \sum_{k=0}^{n} a_k L^{k/\nu''} u^k(w).
	\label{eq:sigma_fitting}
\end{align}
We apply the standard polynomial fitting procedure to those numerical
conductivity data in the DTM phase close to TCP1. The results are shown in Table \ref{tab3} and Fig.~\ref{fig8}(a). The best fitting with expansion order $(n, m)=(1,3)$ gives
\begin{align}
W_\text{TCP} &= 2.67 \pm 0.09,
&
\nu'' &= 1.54 \pm 0.03.
\label{Wc-critical-exponent}
\end{align}

The critical behavior of $\sigma(L,W)$ described by the usual power-law scaling function near TCP1 should cross over to a logarithmic dependence in the DTM phase. In fact, for any finite $W > W_\text{TCP}$, if $L$ is sufficiently large ($L > \xi$), we should see a logarithmic behavior rather than a power law, as we have mentioned in Section~\ref{sec:DOS-metal}. Indeed, already at $W=6$, which is barely inside the DTM phase, the conductivity data can be reasonably fit by a linear function of $\ln L$, see Fig.~\ref{fig8}(b). This suggests that we should only use a relatively narrow range of $W$ to find $\nu''$ from fits to a power law.

\begin{table*}
	\caption{Results of finite-size scaling analysis of the thermal conductivity near TCP1. The conductivity data in the DTM phase are fitted by the polynomial function defined in Eqs.~(\ref{u-w}) and (\ref{eq:sigma_fitting}) with $n=1$ and $m=3,4$. The data in a range of
	$3<W<4$ are fitted. The square brackets are 95\% confidence intervals from 1000 Monte Carlo simulations. The thermal conductivity unit $\sigma_0$ is
	defined by the thermal conductance quantum $G_0 \equiv \pi^2 k_B^2 T/6h$
	as $\sigma_0=G_0/\pi$.}
	\centering
	\begin{ruledtabular}
		\begin{tabular}{cccccccc}
			$L$ &$W$ & $n$& $m$ & GOF & $ W_\text{TCP}$ & $\nu''$ & $\sigma_c/\sigma_0$ \\
			\hline
			24$\sim$120 & 3$\sim$4 &1 & 3 & 0.24 & 2.665[2.461,2.748] & 1.535[1.479,1.599] & 1.926[1.922,1.929] \\
			24$\sim$120 & 3$\sim$4 &1 & 4 & 0.22 & 2.729[2.419,2.814] & 1.536[1.478,1.592] & 1.926[1.923,1.929]\\
		\end{tabular}
	\end{ruledtabular}
\label{tab3}
\end{table*}

\begin{figure}
\centering
\includegraphics[width=3.4in]{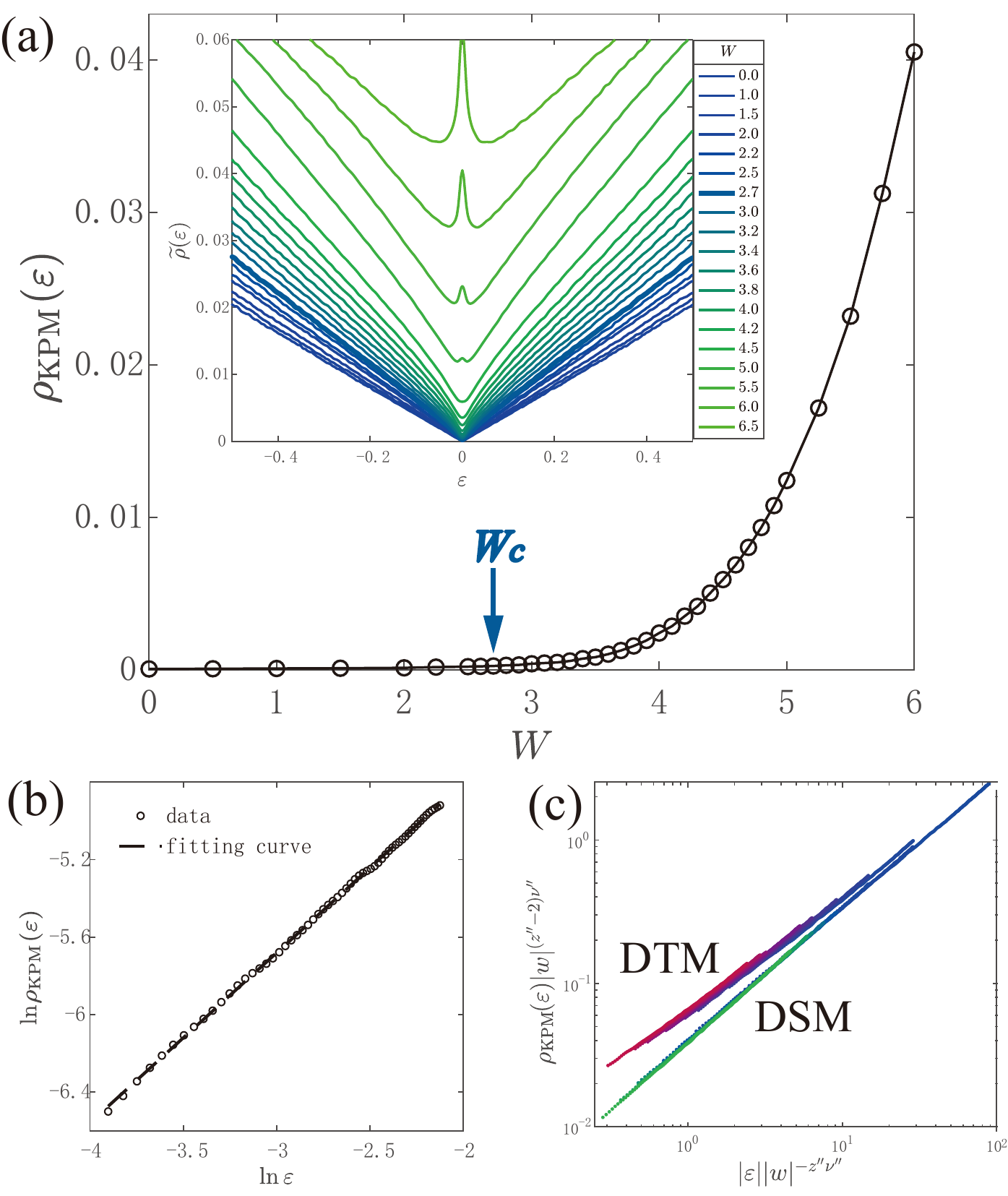}
\caption{(a) $\rho_\mathrm{KPM}(0)$ as a function of the disorder strength $W$ along $\mu=0$. The solid line is a cubic spline interpolation of the data points. Inset: $\rho_\mathrm{KPM}(\varepsilon)$ in a small $\varepsilon$ region for different $W$. (b) $\ln\rho_\mathrm{KPM}(\varepsilon)$ vs. $\ln \varepsilon$ at TCP1 ($W_\text{TCP} = 2.7$). The dashed line is the linear fit $\ln\rho_\mathrm{KPM}(\varepsilon) = a\,\ln\varepsilon +b$ in the range $0.02 \le\varepsilon\le 0.12$. The dynamical exponent $z''$ is extracted from the coefficient $a=(2-z'')/z''$. (c) The scaling collapse of the DOS data near TCP1 along the line of $\mu=0$. We use $z'' = 1.065$, $\nu'' = 1.54$, $W_{\rm TCP} = 2.7$, and $\varepsilon \in [0.05,0.4]$. The data with $W \in [2.9,3.6]$ collapse onto the upper branch that represents the DOS scaling function $f_+$ for the DTM phase. The data with $W\in [1.7,2.6]$ collapse onto the lower branch that represents the DOS scaling function $f_-$ for the Dirac semimetal (DSM) phase.}
\label{fig9}
\end{figure}

The semimetal-metal quantum phase transition can be also characterized by a DOS scaling with a dynamical critical exponent~\cite{Kobayashi2014, Liu16, Syzranov18}. A single-parameter DOS scaling near the zero energy was previously considered for the Dirac semimetal-metal quantum phase transition in three dimensions~\cite{Kobayashi2014}. The same scaling argument applies in two dimensions, and gives the following scaling function for the DOS $\rho(\varepsilon,w)$ near $\varepsilon=0$~\cite{Kobayashi2014, Syzranov18}:
\begin{align}
\rho(\varepsilon,W) &\propto |w|^{(2-z'')\nu''}f_{\pm}(|\varepsilon| |w|^{-z''\nu''}),
\label{dos-scaling}
\end{align}
with the dynamical exponent $z^{\prime\prime}$. $f_{+}(x)$ and $f_{-}(x)$
are universal DOS scaling functions in the DTM ($w>0$) and in the Dirac semimetal ($w<0$).

To evaluate the dynamical exponent $z^{\prime\prime}$ and the scaling functions
$f_{\pm}(x)$, we calculate the DOS along the line $\mu=0$ at different disorder strengths $W$. In the weak disorder region, the numerical value $\rho_{\rm KPM}(0)$ by the kernel polynomial method remains negligibly small ($<10^{-3}$), reflecting the ballistic transport of Dirac-type excitations. The vanishing $\rho_{\rm KPM}(0)$ is consistent with the renormalization group (RG) calculations of Refs.~\cite{Bocquet2000, Mildenberger2007}, which give $\rho(\varepsilon) \propto |\varepsilon|(1 + \alpha W \ln (1/|\varepsilon|))$. Above a certain disorder strength $\rho_{\rm KPM}(0)$ takes a finite value ($>10^{-3}$). The value of $W$ where this happens matches well with the location of TCP1 determined by the conductivity scaling [Fig.~\ref{fig9}(a)]. To evaluate the dynamical exponent $z''$, we fit the low-energy DOS at $W=W_{\rm TCP}=2.7$ by
\begin{align}
\rho(\varepsilon, W=W_\text{TCP}) \propto |\varepsilon|^{(2-z'')/z''},
\end{align}
see Fig.~\ref{fig9}(b). The fitting gives
\begin{align}
z'' = 1.065 \pm 0.0025. 
\label{dynamical-exponent}
\end{align}
We emphasize that the confidence bound for the dynamical exponent is from a single fit, while the true bound must be larger when uncertainties of $W_\text{TCP}$ and the DOS data are included. In Fig.~\ref{fig9}(c)
the DOS data are rescaled according to Eq.~(\ref{dos-scaling}) using the values for $z''$, $\nu''$, and $W_{\rm TCP}$ obtained above. As expected, all data for different $\varepsilon$ and $w<0$ collapse onto one curve and those for $w>0$ collapse onto another curve. These two curves represent the universal DOS scaling functions $f_{\pm}(x)$.


The exponents $\nu''$ and $z''$ at TCP1 are clearly different from those
of the metal-insulator transition points on the DTM-TQH boundary
(point A) and the DTM-AI boundary (point B). This implies that TCP1 is
an unstable fixed point in the low-energy limit, where both $\mu$ and
$w \equiv (W-W_{\rm TCP})/W_{\rm TCP}$ are relevant scaling variables.

We anticipate that the critical behavior near TCP1 is generic, and applies to  other tricritical points in disordered class D models. TCP1 in the isotropic model is described by two copies of the random-mass Dirac fermion, and the lattice Hamiltonian leads to finite couplings between the two. On the other hand, generic tricritical points present in the anisotropic model and TCP2 in the isotropic model, are described by a single copy of the random-mass Dirac fermion. As we have already mentioned, the numerical results of Sec.~V show
that the plateau transitions on the TQH-TQH boundary and on the TQH-AI boundary share the same clean-limit Ising criticality. This indicates that the couplings between the two copies of the Dirac fermion as well as the random mass in each Dirac fermion are irrelevant around the clean-limit fixed point all the way up to TCP1. Therefore TCP1 in the isotropic model can be regarded as two {\it decoupled} generic tricritical points of the class D symmetry. Without any
interference between the two copies, the critical nature of TCP1 must be the same as of a generic tricritical point.

\begin{figure}
\centering
\includegraphics[width=2.8in]{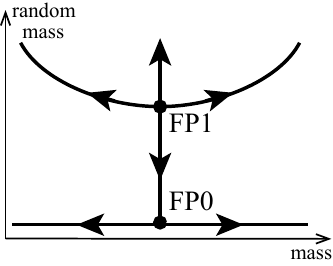}
\caption{Schematic renormalization group (RG) flow around a disorder-induced tricritical point in the 2D symmetry class D. FP0 and FP1 are fixed points in the clean limit and at finite disorder, respectively. The horizontal and vertical axes correspond to $\mu$ and $W$ in the phase diagram of the tight-binding model [see Fig.~\ref{fig1}(a) and (b)].}
\label{fig10}
\end{figure}

Based on all numerical evidence, we provide a schematic renormalization group (RG) flow diagram near a generic tricritical point in class D in two dimensions, see Fig.~\ref{fig10}. The diagram describes both the case of a single disordered Dirac fermion (TCP2) and of two uncoupled copies of disordered Dirac fermions (TCP1). Around the clean-limit fixed point FP0, the uniform mass of the Dirac fermions is relevant while the random mass is marginally irrelevant. The unstable fixed point FP1 exists at finite disorder, corresponding to tricritical points in 2D class D disordered systems. The RG trajectory that starts at FP1 and ends at FP0 corresponds to the plateau transition line between topologically distinct gapped phases. Above FP1, the RG flows are controlled by fixed points of other effective theories, which describe the thermal metal phase and the metal-insulator Anderson transitions.

\section{Conclusions}
\label{sec:conclusions}

In summary, we have provided comprehensive characterizations of phases and quantum phase transitions in a model of the 2D disordered class D topological superconductor. The rich phase diagram comprises three fundamental phases: diffusive thermal metal (DTM), Anderson insulator (AI) and thermal quantum Hall (TQH) phase.

We demonstrated the logarithmic divergence of low-energy DOS in the DTM phase and at the DTM-TQH transition. This implies that the dynamical exponent $z$ at the DTM-TQH transition is 2, the same as the spatial dimension.

By a finite-size scaling analysis of the quasi-1D localization length, we determined the critical 
exponent of the divergent characteristic length 
$\nu = 1.35 \pm 0.04$ for 
the DTM-TQH transition and $\nu = 1.36 \pm 0.05$ for the DTM-AI transition.
Critical conductance distributions at the two transitions are also very similar. Therefore we conclude that the two metal-insulator transitions are controlled by the same fixed point.

\begin{figure*}
\centering
\includegraphics[width=6.5in]{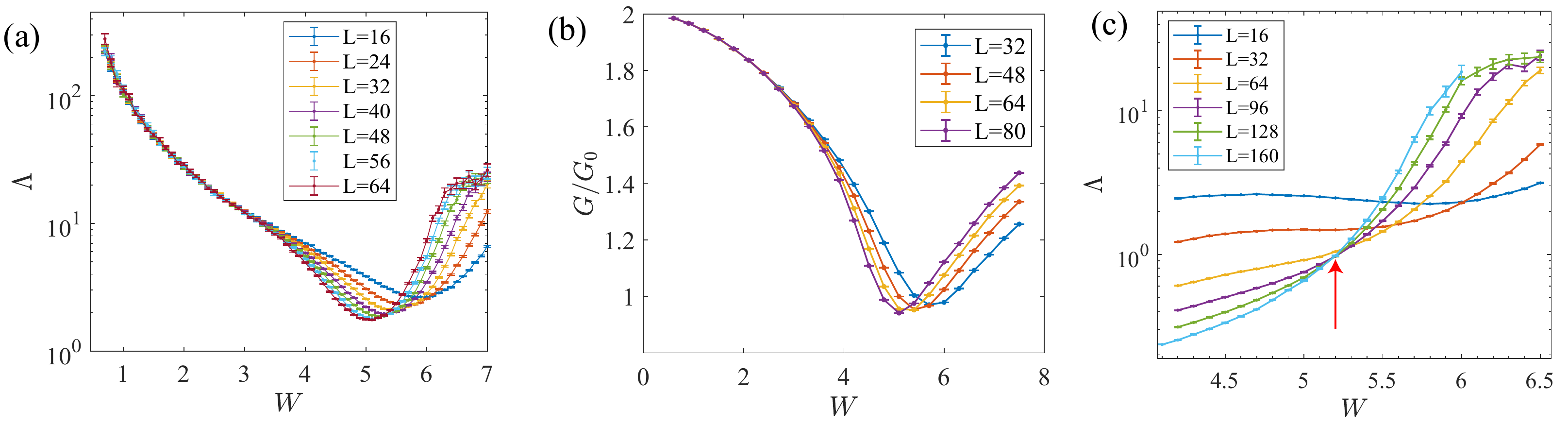}
\caption{(a) Normalized quasi-1D localization length $\Lambda$ and (b) two-terminal conductance $G$ as a function  of the disorder strength $W$ along $\mu=0$. The data points are calculated with the isotropic class D model. $G_0=\pi^2 k_B^2 T/6h$ is the thermal conductance quantum. (c) $\Lambda$ as a function of $W$ at $\mu=0.1$. The red arrow indicates the DTM-TQH transition point at $\mu=0.1$.}
\label{figB1}
\end{figure*}

Thermal quantum Hall (TQH) plateau transitions between distinct topological superconducting phases survive in the presence of disorder, up to tricritical points (TCPs). Around the TQH plateau transition lines, low-energy excitations of the system are described by the field theory of Dirac fermions with random mass. For the isotropic model Eq.~(\ref{bdg}), two TQH transitions degenerate into one along the line of zero chemical potential ($\mu=0$), and two of the tricritical points also merge into one point TCP1. Scaling analysis of the two-terminal conductance shows that the localization length exponent $\nu' = 1$ with high precision,
for both the TQH-TQH and the TQH-AI transitions. This suggests that
the impurity scattering between the two flavors at $\mu = 0$ has no  significant effect on the critical nature of the TQH transition and the tricritical points.

Scaling analyses of the conductivity and the low-energy DOS near TCP1 give the critical exponent of the  
divergent characteristic length $\nu'' \approx 1.54$ 
and dynamical exponent $z'' \approx 1.06$, respectively. 
These values are different from those at the metal-insulator transitions. We deduce from these observations that the tricritical point is an unstable fixed point with two relevant scaling variables on the $\mu-W$ phase plane. This conclusion is in agreement with Ref.~\cite{Medvedyeva2010prb}, but contradicts Ref.~\cite{Kagalovsky2010prb}.

The numerical results reported in this paper are consistent with our recent RG analysis of the 2D Dirac fermions with random mass \cite{pan2021}. The two-loop RG analysis as well as its four-loop extension finds an infra-red unstable fixed point at a
finite disorder strength, where the uniform Dirac mass is a relevant scaling parameter.

The findings strongly suggest that the criticality of DTM-TQH and DTM-AI transitions can be effectively described by a non-linear sigma model,  where the zero-energy DOS near the critical point is non-zero (and even divergent in an infinite system). Meanwhile, the TQH transitions and tricritical points can be described by effective theories of Dirac fermions with vanishing zero-energy DOS~\cite{Syzranov18}.

\begin{acknowledgments}

We thanks Xunlong Luo and Alexander Mirlin for helpful discussions, and Bjorn Sbierski 
for critical reading of the manuscript. T.W., Z.P. and R.S. were supported by the 
National Basic Research Programs of China (No. 2019YFA0308401) and the National Natural 
Science Foundation of China (Grant No. is 11674011 and 12074008). T.O. was supported 
by JSPS KAKENHI Grants No. 16H06345 and 19H00658. 

\end{acknowledgments}

\appendix

\section{Tight-binding model of a 2D class D topological insulator}
\label{appendix-A}

The diagonalization of the matrix $\mathbb{H}$ in Eq.~\eqref{bdg-matrix} can be restated as the solution of a tight-binding model of a topological insulator with two orbitals ($a$,$b$) per site:
\begin{align}
	\mathcal{H} = &\sum_{\bm j} (\varepsilon_{\bm j} +\mu)
(a_{\bm j}^\dagger a_{\bm j}^{\vphantom{\dagger}}
- b_{\bm j}^\dagger b_{\bm j}^{\vphantom{\dagger}})
+ \Delta \!\ \sum_{\bm j} \big[\mathrm{i} (a_{\bm j + \bm{e}_x}^\dagger b_{\bm j}^{\vphantom{\dagger}}
\nonumber \\
& + b_{\bm j+\bm{e}_x}^\dagger a_{\bm j}^{\vphantom{\dagger}})
+ (a_{\bm j+\bm{e}_y}^\dagger b_{\bm j}^{\vphantom{\dagger}}
- b_{\bm j+\bm{e}_y}^\dagger a_{\bm j}^{\vphantom{\dagger}}) + \mathrm{h.c.} \big]
\nonumber \\
& +\!\ \sum_{\bm j} \sum_{\nu=x,y} t_{\nu} \big[
	(a_{\bm j+\bm{e}_\nu}^\dagger  a_{\bm j}^{\vphantom{\dagger}} -
	b_{\bm j+\bm{e}_\nu}^\dagger b_{\bm j}^{\vphantom{\dagger}})+ \mathrm{h.c.} \big].
	\label{eq:A}
\end{align}
In this picture, the two orbitals represent particles and holes of the original BdG Hamiltonian~\eqref{bdg}, and the inter-orbital hoppings correspond to the $p$-wave pairing amplitudes $\Delta$. We use this topological model for the transfer matrix calculations of the localization length and the conductance as well as the kernel polynomial method calculation of the DOS. The conductance and the DOS calculated in the tight binding model Eq.~\eqref{eq:A} are those of the
Bogoliubov quasiparticles in class D disordered superconductor model Eq.~\eqref{bdg}.

To calculate the two terminal
Landauer conductance of the disordered topological insulator model, we attach two leads to 
the disordered model. Each lead consists of decoupled 1D metallic wires, 
\begin{align}
    \mathcal{H}_{\rm lead} = t_{\rm lead}\sum_{\bm j}\left[ a_{\bm{j}+\bm{e}_x}^\dagger a_{\bm j}^{\vphantom{\dagger}} + b_{\bm{j}+\bm{e}_x}^\dagger b_{\bm j}^{\vphantom{\dagger}} + {\rm h.c}\right],  
\end{align}
with a hopping amplitude $t_{\rm lead}$. 
The leads are semi-infinite and disorder-free, and the plane wave eigenstates 
of the leads are labeled by wave vector $k_x$, where the chemical potential is equal to
$2 t_{\rm lead}\cos(k_x)$.  In the present paper,
we set the chemical potential at the half-filling point ($k_x=\pi/2$) and 
calculate the Landauer conductance along $x$.

\begin{figure*}
\centering
\includegraphics[width=7in]{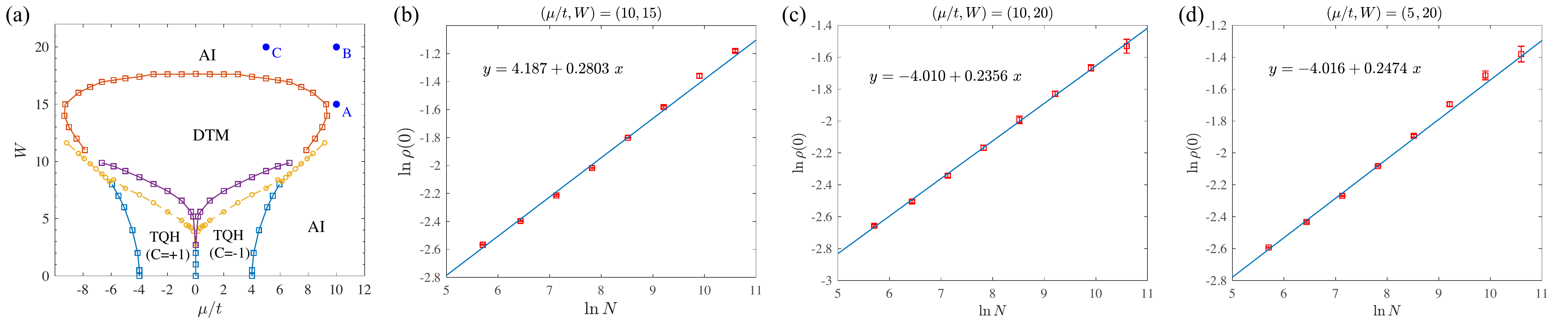}
\caption{ (a) Phase diagram of the isotropic class D model taken from FIG.~\ref{fig1}(b). (b)-(d) show the zero-energy DOS $\rho_{\rm KPM}(0)$ as a function of an expansion order $N$ of kernel polynomial method at three parameter points in the Anderson insulator phase. The locations of the parameter points are specified by blue dots in the panel (a) as A, B, and C, for the data shown in panels (b), (c), and (d), respectively.
The calculation is carried out with the square-geometry sample ($L\times L$) of size $L=1000$ with periodic boundary condition in both $x$ and $y$ directions. The expansion order $N$ ranges from 300 to 4000. The error bar is the standard deviation of 4 samples.}
\label{appendix-fig2}
\end{figure*}

\section{Finite-size effects near the tricritical point TCP1}
\label{appendix-B}

Both the normalized quasi-1D localization length $\Lambda$ and the two-terminal conductance $G$ should exhibit scale-invariant behavior along the plateau transition line between topologically distinct gapped phases below the tricritical point TCP1. When the system undergoes the semimetal-metal quantum phase transition and enters the diffusive thermal metal (DTM) phase, both $\Lambda$ and $G$ are expected to increase monotonically with the system size $L$. Figures~\ref{figB1}(a) and~\ref{figB1}(b) show, respectively, $\Lambda$ and $G$ as functions of the disorder strength $W$ along the line $\mu=0$, for different system sizes. Notice that in a weak disorder region ($0<W < 2.5$), both quantities indeed take scale-invariant critical values. The critical value of $\Lambda$ diverges in the clean limit, while in the same limit, the critical value of $G$ converges to an integer in units of $G_0$.

However, above TCP1 at $W_\text{TCP} \simeq 2.7$, the metallic nature of the DTM phase is manifest only for sufficiently large $W \gtrsim 6$, where both $\Lambda$ and $G$ increase with $L$. In the range $5 < W < 6$, we observe a non-monotonic behavior in both $\Lambda$ and $G$: they decrease with $L$ for smaller $L$ values and increase with $L$ for larger $L$ values. When $W$ gets even closer to $W_{\rm TCP}$ ($2.7< W < 5$), both $\Lambda$ and $G$ decrease with $L$ up to the largest numerically available system size.

We attribute the non-monotonic $L$-dependence of $\Lambda$ and $G$ to
finite-size effects near the tricritical point. As is typical in the presence of such effects, above $W_\text{TCP}$, curves for two successive values of $L$ cross at a point, but this point systematically shifts toward smaller $W$ upon increasing $L$. We expect that for sufficiently large $L$, the crossing points will finally collapse to the tricritical point TCP1 at $W = W_\text{TCP}$. In view of the finite-size effects, we have to either simulate very large systems, or resort to other methods to precisely determine the position of TCP1. In the main text, we used the conductivity scaling analysis to find the critical disorder for the semimetal-metal transition along $\mu=0$. 

Similar finite-size effects are also observed close to TCP1 along $\mu \ne 0$, as shown in Fig.~\ref{figB1}(c). A scale-invariant point for $\Lambda(L)$ for smaller $\mu$ can be found only for $L \ge 96$. We interpret the value of $W$ at the crossing point of the $L=96, 128, 160$ curves in Fig.~\ref{figB1}(c) as the critical disorder strength $W_{c}$ of the DTM-TQH transition at $\mu = 0.1$. The DTM-TQH transition line determined in this way shows a sharp `dip' structure near the TCP1 at $(W_\text{TCP}, \mu) = (2.7, 0)$. Farther away from the TCP1, the finite-size effects of
$\Lambda(L)$ are weaker, and the scale-invariant point of $\Lambda(L)$ (at the metal-insulator transition) can be observed at smaller $L$. Precise determination of the MIT lines requires a polynomial fitting procedure, while a rough estimate from the plots of $\Lambda(L)$ vs. $W$ is
enough to reveal the structure of the phase diagram.

\section{DOS in the Anderson insulator phase near the MIT transition line}
\label{appendix-D}

The DOS in the localized phase near the MIT point shows weakly singular structures around zero energy, as seen in Fig.~\ref{fig2}. In this paper, we calculate the DOS by the kernel polynomial expansion method~\cite{KPM2006},
where the $\delta$-function is approximated by its finite-order expansion in terms of the Chebyshev polynomials. Due to this approximation, the energy resolution is limited by the truncation order of the polynomial $N$. As a consequence, the numerical DOS $\rho_{\rm KPM}(\varepsilon)$ is different from the true DOS $\rho(\varepsilon)$ at finite $N$:
\begin{align}
    \rho(\varepsilon) & \equiv \frac{1}{V}\sum_{i} \delta(\varepsilon-\varepsilon_{i}), \nonumber \\
    \rho_{\rm KPM}(\varepsilon) & \equiv \frac{1}{V}\sum_{i} \frac{\pi^{-1} a N^{-1}}{(\varepsilon-\varepsilon_i)^2+(a N^{-1})^2}.
\end{align}
Here $a$ is a coefficient of order unity. The true DOS
and the numerical DOS are related by the energy integral,
\begin{align}
    \rho_{\rm KPM}(\varepsilon) = \int^{+\infty}_{-\infty} dx \rho(x) \frac{\pi^{-1}
    a  N^{-1}}{(\varepsilon-x)^2+(a N^{-1})^2}. \label{eq:kpm-true}
\end{align}
The relation leads to
\begin{align}
    \rho_{\rm KPM}(0) \propto
    \begin{cases}
      \ln N, & \mbox{if } \rho(\varepsilon) \propto \ln \big(1/|\varepsilon|\big), \\
      N^{\alpha}, & \mbox{if } \rho(\varepsilon) \propto |\varepsilon|^{-\alpha},
    \end{cases}
\end{align}
with $\alpha>0$. Namely, when the true DOS shows the logarithmic (power-law) divergence in $\varepsilon$, the numerical DOS at the zero-energy shows the logarithmic (power-law) divergence in $N$. The 
same conclusion is drawn numerically, when $\rho_{\rm KPM}(\varepsilon)$ in Eq.~(\ref{eq:kpm-true}) 
is given by an integral of Gaussian kernel.

To study the singularity of $\rho(0)$ in the localized phase near the MIT line, we calculate the DOS by the kernel polynomial method with different values of the truncation order $N$ at several parameter points in the localized phase (Fig.~\ref{appendix-fig2}). The numerical results show that a linear fitting works well in the plot of $\ln \rho_{\rm KPM}(0)$ vs. $\ln N$, where the linear coefficient is small and slightly differs for different points in the phase diagram. The results suggest that the low-energy DOS shows a power-law divergence $\rho(\varepsilon) \sim |\varepsilon|^{-\alpha}$, with a non-universal exponent $\alpha$. The power-law divergence is consistent with
the Griffiths effects previously proposed in a study of a 2D network model in  class D~\cite{Mildenberger-Griffiths-2006}.

\section{Finite-size scaling of conductance in the DTM phase}
\label{appendix-C}

\begin{figure}
\centering
\includegraphics[width=3.3in]{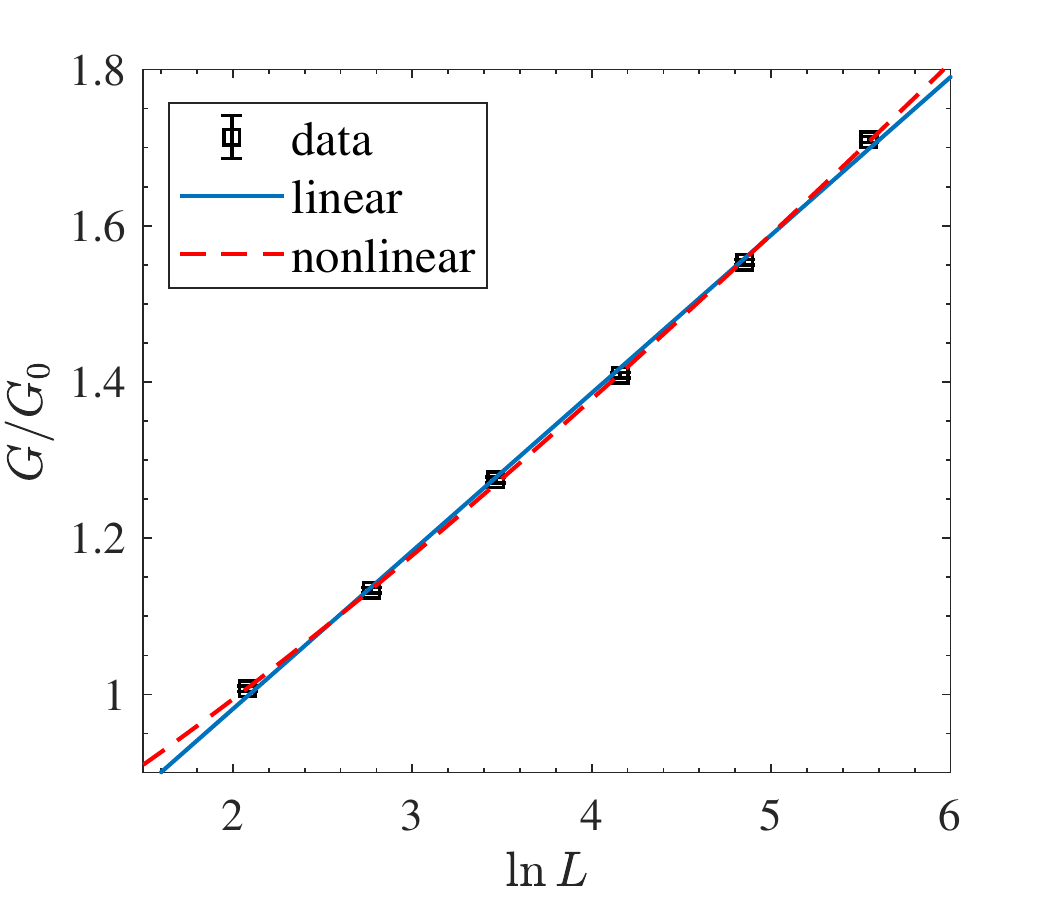}
\caption{Conductance as a function of $\ln L$ in the DTM phase at $(\mu/t,W) = (1,10)$. The two-terminal conductance of the square geometry, $L \times L$, is calculated with the periodic boundary condition along the transverse direction. $L$ ranges from 8 to 256. The error bar is the standard error of $10^4$ samples. 
The blue straight line is a linear fit to Eq.~\eqref{g-0}, resulting in $a_0 = 0.202 \pm 0.003$. The red dashed line is a non-linear fit to Eq.~\eqref{gL}, resulting in $a_0 = 0.28\pm 0.02$, $a_1 = -0.13 \pm 0.03$.}
\label{appendix-fig3}
\end{figure}

The metallic phase in class D is amenable to analytical treatment, since it is described by a weakly-coupled 2D sigma model~\cite{Evers2008RMP, Senthil2000prb, Bocquet2000}. A perturbative analysis of the sigma model leads to
the beta function 
\begin{align}
\beta(g) \equiv \frac{dg}{d\ln L} = a_0 + \frac{a_1}{g} + \frac{a_2}{g^2} + \frac{a_3}{g^3} + \ldots,
\label{beta-g}
\end{align}
where the coefficients $a_i$ can be read off from Refs.~\cite{wegner89, Evers2008RMP}. These and other references give beta functions in terms of the coupling constants $t$ of various sigma models, and one has to be careful about the relation between $g$ and $t$. The first two coefficients are
\begin{align}
a_0 &= \frac{1}{\pi}, & a_1 &= - \frac{2}{\pi^2}.
\label{ai-analytical}
\end{align}

For sufficiently large $L$ it is sufficient to keep only the leading term ($a_0$) in the beta function, which then leads to a logarithmic divergence of the conductance with respect to the system size $L$:
\begin{align}
    g^{(0)}(L) = g_0 + a_0 \ln L,
    \label{g-0}
\end{align}
similar to the weak anti-localization in the 2D sigma model in the symplectic class AII~\cite{wegner89, Asada-Quantum-2006}. 

To test this logarithmic scaling, we compute the Landauer conductance $G$ in the square geometry ($L\times L$) at a point in the DTM phase at $(\mu/t,W)=(1,10)$ for different system sizes $L$ (Fig.~\ref{appendix-fig3}). A logarithmic function $g = g_0 + a_0 \ln L$ gives a reasonable fit in the range $L \in [8,258]$, though the coefficient $a_0 \approx 0.20$ differs from the sigma model prediction $a_0 = 1/\pi \approx 0.32$.
We attribute this discrepancy to insufficient systems sizes $L$. In the range of $L$ available to us, the corrections coming from the higher order
terms in the beta function may not be negligible.

Let us consider Eq.~\eqref{beta-g} keeping $a_0 + a_1/g$ in the right-hand side. This equation can be solved exactly, and $g(L)$ can be expressed in terms of the Lambert $W$ function. 
For our purposes 
it is sufficient to solve the equation iteratively. We use the function~\eqref{g-0} as the zeroth approximation, and substitute it into the equation:
\begin{align}
\frac{dg}{dl} &= a_0 + \frac{a_1}{g_0 + a_0 l}, 
&
l &\equiv \ln L.
\end{align}
This is easily solved:
\begin{align}
g(L) = g_0 + a_0 \ln L + \frac{a_1}{a_0} \ln\Big(1 + \frac{a_0}{g_0} \ln L\Big). 
\label{gL}
\end{align}
Using this function to fit the data we obtain the red dashed line in Fig.~\ref{appendix-fig3}, with the fitting parameters
\begin{align}
a_0 & =  0.28\pm 0.02, & a_1 & = -0.13 \pm 0.03,
\end{align}
in reasonable agreement with the analytical values~\eqref{ai-analytical}. The value of the 
coefficient $a_0$ is close to the one numerically 
obtained in Ref.~\cite{Fulga-Temperature-2020}.

\bibliography{2dDRef}
\end{document}